\documentclass[10pt]{article}
\usepackage{amsmath}
\usepackage{cite} 
\usepackage{indentfirst}
\usepackage{latexsym}
\usepackage{pstricks}
\usepackage{graphicx}

\long\def\ca#1\cb{} 

\newcommand{\ad}{^\dagger }

\newcommand{\AND}{\mbox{\small AND}}

\newcommand{\becs}{\begin{cases}}
\newcommand{\bem}{\begin{matrix}}

\newcommand{\dg}{^\circ } 
\newcommand{\dya}[1]{|#1\rangle\langle#1|}
\newcommand{\dyad}[2]{|#1\rangle\langle#2|}

\newcommand{\encs}{\end{cases}}
\newcommand{\enm}{\end{matrix}}

\newcommand{\hf}{{\textstyle\frac{1}{2} }}

\newcommand{\inp}[1]{\langle#1|#1\rangle }
\newcommand{\inpd}[2]{\langle#1|#2\rangle }
\newcommand{\ket}[1]{|#1\rangle }

\newcommand{\lra}{\leftrightarrow }

\newcommand{\mat}[1]{\left(\begin{matrix}#1\end{matrix}\right)}

\newcommand{\mte}[2]{\langle#1|#2|#1\rangle }
\newcommand{\mted}[3]{\langle#1|#2|#3\rangle }

\newcommand{\NOT}{\mbox{\small NOT}}
\newcommand{\od}{\odot }

\newcommand{\ot}{\otimes }

\newcommand{\ra}{\rightarrow }
\newcommand{\Ra}{\Rightarrow }

\newcommand{\st}{\sqrt{2}}
\newcommand{\tm}{\times }
\newcommand{\Tr}{{\rm Tr}}

\newcommand{\vbl}{\,\boldsymbol{|}\,}

\newcommand{\HC}{{\mathcal H}}

\newcommand{\al}{\alpha }
\newcommand{\bt}{\beta }
\newcommand{\gm}{\gamma }
\newcommand{\Gm}{\Gamma }
\newcommand{\dl}{\delta }

\newcommand{\ep}{\epsilon}

\newcommand{\zt}{\zeta }

\newcommand{\lm}{\lambda }

\newcommand{\om}{\omega }
\newcommand{\Om}{\Omega }

\def\outl#1{\par{\medskip\noindent\hspace*{0.1cm}\bf
      \mathversion{bold}#1\mathversion{normal}\smallskip} }
   \def\xa{} \def\xb{}  

 \def\outl#1{}\def\xa{}\def\xb{}

\ca
 \def\outl#1{\par{\medskip\noindent\hspace*{.5cm}\bf
      \mathversion{bold}#1\mathversion{normal}\smallskip} }
 \long\def\xa#1\xb{} 
  
\cb

\begin{document}

\title{What Quantum Measurements Measure}
\author{Robert B. Griffiths
\thanks{Electronic mail: rgrif@cmu.edu}\\ 
Department of Physics,
Carnegie-Mellon University,\\
Pittsburgh, PA 15213, USA}
\date{Version of 13 September 2017}
\maketitle  


\vspace{-7 ex}

\xa
\begin{abstract}
  A solution to the second measurement problem, determining what prior
  microscopic properties can be inferred from measurement outcomes
  (``pointer positions''), is worked out for projective and generalized (POVM)
  measurements, using consistent histories. The result supports the idea that
  equipment properly designed and calibrated reveals the properties it was
  designed to measure. Applications include Einstein's hemisphere and Wheeler's
  delayed choice paradoxes, and a method for analyzing weak measurements
  without recourse to weak values. Quantum measurements are noncontextual in
  the original sense employed by Bell and Mermin: if $[A,B]=[A,C]=0,\,
  [B,C]\neq 0$, the outcome of an $A$ measurement does not depend on whether it
  is measured with $B$ or with $C$. An application to Bohm's model of the
  Einstein-Podolsky-Rosen situation suggests that a faulty understanding of
  quantum measurements is at the root of this paradox.
\end{abstract}
\xb

\tableofcontents
\xa
\xb

	\section{Introduction}
\label{sct1}
\xa


\subsection{The Second Measurement Problem \label{sbct1.1}}

\xb
\outl{Measurement problem in SQM}
\xa

The \emph{measurement problem} is a central issue in quantum foundations,
because textbook quantum mechanics uses the idea of a measurement to give a
physical interpretation to probabilities generated from a quantum wavefunction,
but never explains the measurement process itself in terms of more fundamental
quantum principles. If, as is widely believed, quantum mechanics applies to
macroscopic as well as microscopic phenomena, then it should be possible, at
least in principle, to describe actual laboratory measurements in terms of
basic quantum properties and processes, rather than employing ``measurement''
as an unanalyzed primitive.

\xb
\outl{First and second measurement problems defined}
\xa

It is convenient to divide the measurement problem into two parts. The
\emph{first measurement problem}, which is at the center of most discussions in
the literature, is to understand how the measurement process can result in a
well-defined macroscopic outcome or \emph{pointer position}, to use the archaic
but picturesque language of the foundations community, rather than some strange
quantum superposition of the pointer in different positions, as results in many
cases from a straightforward application of unitary time development:
Schr\"odinger's equation leads to Schr\"odinger's cat.
But even if the first measurement problem is solved, so the pointer comes to
rest at a single position, the \emph{second measurement problem} remains: what
can one infer from the pointer position regarding the microscopic situation
that existed before the measurement took place, which the apparatus was
designed to measure? Experimental physicists talk all the time about gamma rays
triggering a detector, neutrinos arriving from the sun, and other microscopic
objects or events which are invisible, and whose existence can only be inferred
from the macroscopic outcomes of suitable measurements. Should we take this
talk seriously? Maybe we do, but why, if the second measurement problem remains
unresolved? Would we have any confidence in the stories told us by cosmologists
if they did not understand the operation of their telescopes well enough to
interpret the data these instruments provide?

\xb
\outl{Nested MZ, ctfl communication $\Ra$ measurement problem
  hinders research, teaching}
\xa

A recent (and at the time of writing continuing) controversy
\cite{Vdmn13,Grff16} about the path followed by a photon passing through an
interferometer on its way to a detector shows how difficult it is to analyze,
using the tools of textbook quantum theory, with perhaps some additional ad hoc
principles, a microscopic situation that is really not very complicated. This
problem is, in turn, related to a hotly contested claim, published in a
reputable journal, that information can be sent between two parties by means of
a photon that is actually never---or at least hardly ever---present in the
optical fiber that connects them \cite{LAAZ15,Vdmn16,LAAZ16}.
What this suggests is that the failure of quantum physicists to solve the
measurement problem(s) is not only an intellectual embarrassment---surely it is
that, as pointed out by some leading physicists (see \cite{Gsn17} and Sec.~3.7
of \cite{Wnbr15})---but also a serious impediment to ongoing research in areas
such as quantum information, where understanding microscopic quantum properties
and how they depend on time is central to the enterprise. In addition, a fuzzy
understanding of quantum principles makes the subject hard to teach as well as
to learn. Students confused by unfamiliar mathematics are not helped by the
absence of a clear physical interpretation of what the mathematics means,
something which neither textbooks nor instructors seem able to provide.

\xb
\outl{Measurement problem addressed using CH. Advantages of CH}
\xa

In this paper the second (and, incidentally, the first) measurement problem is
addressed using the consistent histories, also known as decoherent
histories, interpretation of quantum mechanic. While this approach is
controversial (as is everything else in quantum foundations) it possesses
specific principles and clear rules for applying and interpreting quantum
theory at the microscopic level. These principles are comparatively few in
number, include no reference to measurements, and apply universally to all
quantum processes, whether microscopic to macroscopic, ``from the quarks to the
quasars.'' They are, so far as is known at present, consistent in the sense
that when properly applied they do not lead to contradictions, and they have
resolved (perhaps 'tamed' would be a better term) various quantum paradoxes;
see Chs.~21 to 25 of \cite{Grff02c} for a number of examples.

\subsection{Article Overview \label{sbct1.2} }

\xb
\outl{Sec. 2: Einstein, Wheeler paradoxes. Sec. 3: CH}
\xa

The remainder of the paper is structured as follows. Section~\ref{sct2}
explores the second measurement problem from a phenomenological perspective
using two paradoxes, the first by Einstein and the second by Wheeler, that show
why the problem is both difficult and confusing. Section~\ref{sbct2.3} is a
brief discussion of how a measurement apparatus can be calibrated to ensure its
reliability. Next a brief summary of the consistent histories approach, along
with references to literature that provides further details, constitutes
Section~\ref{sct3}; readers already familiar with consistent histories ideas
can skip it.

\xb
\outl{Sec. 4: Projective; POVMs; backwards map; nondestructive; preparations;
density ops}
\xa

Section~\ref{sct4} is the heart of the paper, and contains the key ideas needed
to address the second measurement problem both for projective measurements,
Sec.~\ref{sbct4.1}, and for generalized measurements (positive operator-valued
measures, or POVMs),
Sec.~\ref{sbct4.2}. The emphasis is on simple cases of single measurements;
situations where there are several successive measurement on the same system
are not discussed, though the histories methodology can also be extended to
such situations. A useful conceptual tool, which so far as we know has not been
pointed out previously, is the backwards map from output (pointer) states to
earlier microscopic properties. It is very helpful in identifying the
microscopic properties which have been measured in the case of a generalized
measurement.
A separate Sec.~\ref{sbct4.3} discusses nondestructive measurements and
preparations, both closely related to von Neumann's measurement model. This may
assist the reader in connecting the approach followed in this paper to ideas,
such as wavefunction collapse, frequently encountered in textbook treatments
and quantum foundations literature. The final Sec.~\ref{sbct4.4} has a few
comments about density operators.

\xb
\outl{Sec.~\ref{sct5}: applications: two $\lra$ Sec.~\ref{sct2}; 
POVM; weak measurement; noncontextual QM; EPRB}
\xa

Next in Sec.~\ref{sct5} the tools developed in Sec.~\ref{sct4} are applied to
six different situations, where the first two, Secs.~\ref{sbct5.1} and
~\ref{sbct5.2}, are closely related to the examples discussed earlier in
Sec.~\ref{sct2}.  The third, Sec.~\ref{sbct5.3}, is an elementary but not
entirely trivial example of a POVM that is not a projective measurement. A
fairly elementary, but again nontrivial, example in Sec.~\ref{sbct5.4} shows
how a weak measurement can be interpreted in terms of quantum properties
instead of the widely used ``weak values.''
The last two applications address topics which often come up in the quantum
foundations literature, and are hence somewhat controversial. It is argued in
Sec.~\ref{sbct5.5}, using a less formal and more physical approach than
\cite{Grff13b}, that if one uses Bell's original definition of ``contextual,''
quantum mechanics is in fact noncontextual, despite confusing claims to
the contrary. Finally, the Bohm (spin singlet) model of the famous
Einstein-Podolsky-Rosen paradox is discussed in Sec.~\ref{sbct5.6} from the
perspective of what one can infer from measurements on one of the spin-half
particles about its prior properties and those of the other spin-half particle.

\xb
\outl{Sec. 6: Summarizes conclusions}
\xa

The final Sec.~\ref{sct6} is an attempt to summarize the most important
conclusions about what it is that quantum measurements measure, while
summarizing the principles which make it possible for the consistent histories
interpretation to arrive at a satisfactory resolution of the second (as well as
the first) measurement problem.

\subsection{Notation and Acronyms \label{sbct1.3} }

In addition to standard Dirac notation note the following:\\
${\ }\bullet$\ The symbol $\od$ is a tensor product symbol equivalent to the
usual $\ot$, but used in a quantum history to separate an earlier event to the
left of $\od$ from a later event to
its right.\\
${\ }\bullet$\ The curly brackets in $[\Psi_0]\od\{A,B\}\od C$ indicate two
histories: $[\Psi_0]\od A\od C$
and  $[\Psi_0]\od B\od C$.\\
${\ }\bullet$\ $[\psi] = \dya{\psi}$ when $\ket{\psi}$ is a normalized state.
Square brackets may be omitted if the meaning is clear: $\Pr(z_1^+)$ in place
of $\Pr([z^+]_1)$ \\
${\ }\bullet$\ Superscripts are used as \emph{labels} and not as exponents on
projectors (where an exponent is never needed) and sometimes on other symbols
in order to reserve the subscript position to label system or the time. Thus
$z^+_{a1}$ and $M_2^-$ in $\Pr(z^+_{a1},M^-_2)$ refer to $[z^+]$ for system $a$
at time $t_1$, and $M^-$ at time $t_2$.

The following acronyms (to be precise, initialisms) are placed here for ready
reference:\\
${\ }\bullet$\ EPR = Einstein-Podolsky-Rosen\\
${\ }\bullet$\ PDI = projective decomposition of the identity. See \eqref{eqn6}
in
Sec.~\ref{sbct3.2}\\
${\ }\bullet$\ POVM = positive operator-valued measure. See Sec.~\ref{sbct4.2}\\

\xb
\section{Measurement Phenomenology}
\label{sct2}
\xa

\xb
\subsection{Einstein's Paradox \label{sbct2.1}}
\xa

\begin{figure}[h]
$$\includegraphics{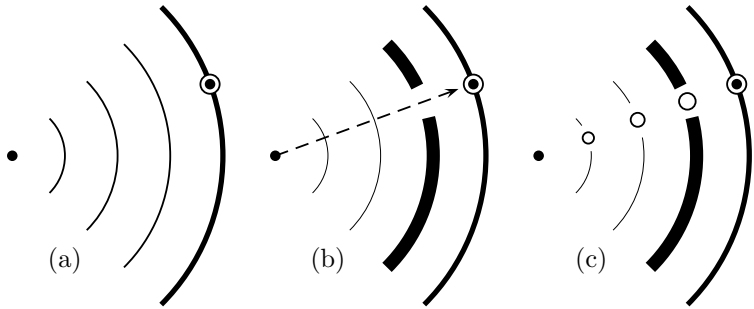}$$
\caption{%
Einstein paradox. (a) Spherical wave; (b) Particle moving on straight line
through collimator; (c) Quantum wavepacket passing through collimator.
}
\label{fgr1}
\end{figure}

\begin{figure}[h]
$$\includegraphics{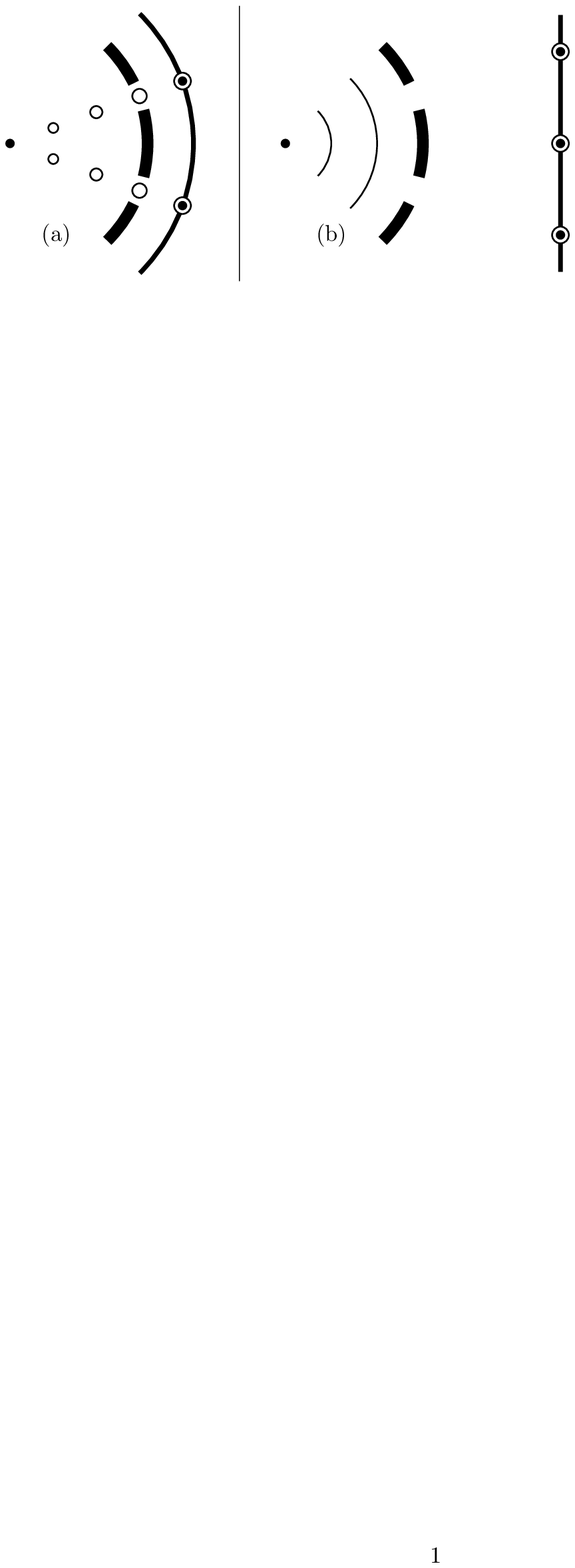}$$
\caption{%
  (a) Collimator with two holes. (b) Fluorescent screen a large distance to the
  right of the collimator. Due to constructive interference of waves coming
  from the two holes a particle can sometimes be observed in a region which is
  classically forbidden. }
\label{fgr2}
\end{figure}

\xb
\outl{Spherical wave impinging on spherical screen}
\xa

Figure ~\ref{fgr1}(a) shows Einstein's paradox (pp.~115-117 in \cite{Jmmr74},
pp.~440-442 in \cite{BcVl09}). A particle emerges from a small hole at the left
and propagates as a spherical wave towards a curved fluorescent screen where
its arrival is signalled by a flash of light at a particular point on the
screen, a point which varies randomly on successive repetitions of the
experiment. It seems as if the quantum wave collapses instantly when the
particle reaches the screen, a result which bothered Einstein as it would mean
a superluminal effect if every point on the screen is equidistant from the
hole. An experimental physicist, on the other hand, might say that the particle
travels on a straight line from the source to the screen, and could support
that explanation by placing a collimator, a thick plate with a hole in it,
between the source and the screen, and noting that now flashes are detected
only at places on the screen which are connected to the source by a straight
line passing through the hole, Fig.~\ref{fgr1}(b).

\xb
\outl{Quantum wavepacket passing through collimator hole}
\xa

But isn't this second perspective classical, not quantum mechanical? No, for
there is a good quantum mechanical description in which the particle is a small
wave packet traveling from the source to the screen, Fig.~\ref{fgr1}(c); one
only has to assume that the particle emerging from the source is described by
such a wave packet whose initial direction of propagation is random from one
run to the next. (And this gets around another problem with wavefunction
collapse. If the particle reaches the screen, does this mean that its failing
to interact with the collimator has collapsed the spherical wave enough so that
it can fit through the hole?) 

\xb 
\outl{Collimator with two holes producing  interference} 
\xa

Continuing on, if the collimator has two holes, Fig.~\ref{fgr2}(a), one will
observe flashes on the screen due to particles which have passed through one
hole or the other, but never simultaneous flashes behind both holes. Again,
easy to understand using the picture of little wave packets. But consider the
situation in Fig.~\ref{fgr2}(b) where, if the two holes are formed very
carefully and the fluorescent screen placed a long distance away, the result
will be an interference pattern with the distance between fringes determined
by, among other things, the distance between the two holes and the de Broglie
wavelength of the quantum particle. The particle must, in this case, be thought
of as a wave passing simultaneously through both holes and emerging behind them
with a well-defined phase. We have arrived at the double slit or two hole
paradox so well described by Feynman \cite{FyLS651}.

\xb
\outl{Multiple quantum descriptions}
\xa

Everyone knows that quantum particles are waves, and quantum waves are
particles. The gedanken experiments just discussed, especially the contrast
between Fig.~\ref{fgr2}(a) and (b), illustrate the fact that sometimes a
particle (fairly well localized wavepacket) and sometimes a wave (coherence in
phase over a macroscopic distance) description is needed in order to understand
what is going on. The need to use different, and seemingly incompatible,
descriptions is one of the fundamental difficulties behind the second measuring
problem. One aim of the present article is to show how it can be
addressed without invoking \emph{retrocausation}: a future measurement 
influencing past behavior.

\xb
\subsection{Mach-Zehnder with Removable Beamsplitter \label{sbct2.2}} 
\xa

\begin{figure}[h]
$$\includegraphics{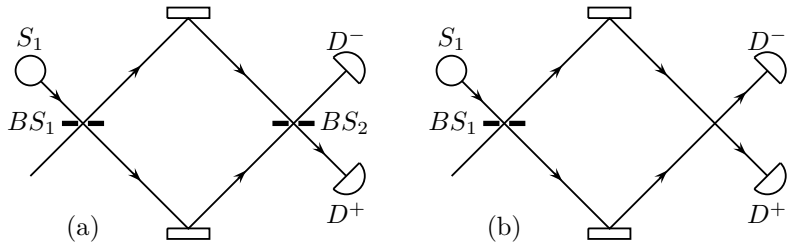}$$
\caption{%
  Mach-Zehnder interferometer (a) with a source $S_1$, two beamsplitters $BS_1$
  and $BS_2$, and detectors $D^+$ and $D^-$; (b) with the second beamsplitter
  removed. }
\label{fgr3}
\end{figure}

\xb
\outl{Mach-Zehnder with 2d beamsplitter in place}
\xa

Einstein's paradox becomes easier to analyze if we consider the case of a
Mach-Zehnder interferometer, Fig.~\ref{fgr3}(a), with an upper and lower arm
connecting two beam splitters $BS_1$ and $BS_2$, and the phases adjusted so
that a photon---hereafter referred to as a `particle'---from the source $S_1$
on the left is always detected by the lower detector $D^+$ on the right.
That the particle is, in some sense at least, in both the upper and the lower
arm while inside the interferometer can be checked by inserting two phase
shifters, one in each arm. One then finds that, depending on the choice of
phases, the particle will sometimes be detected in $D^+$ and sometimes in
$D^-$. However, if both phases are identical, the particle will always be
detected in $D^+$. Additional checks can be made by blocking either the upper
arm or the lower arm, and noting that when one arm is blocked the particle
will sometimes arrive in $D^+$ and sometimes in $D^-$. 

\xb
\outl{Particle path thru MZ with 2d beamsplitter removed}
\xa

If, on the other hand, the second beam splitter is absent, Fig.~\ref{fgr3}(b),
the experimentalist will say that a particle detected in $D^+$ was originally
in the upper arm of the interferometer, and if detected in $D^-$ it was in the
lower arm, as these are the direct paths from the first beam splitter to the
detectors. This can be checked by placing barriers in the upper or lower arms
of the interferometer and noting that a barrier in the upper arm will prevent
the particle arriving at $D^+$, and one in the lower arm suppresses counts in
$D^-$. Similarly, if a nondestructive measuring device, something which will
register the particle's presence without seriously perturbing its motion, is
placed in one of the arms, its outcome will show the expected correlation with
the final detectors.

\xb
\outl{Wheeler's delayed choice; retrocausation?}
\xa

Wheeler's delayed choice paradox \cite{Whlr78} comes from asking what will
occur if just before the particle arrives at $BS_2$, when it has already passed
$BS_1$ and is inside the interferometer, the second beamsplitter is removed.
Alternatively, suppose that the second beamsplitter is absent while the
particle is traversing the interferometer, but is suddenly inserted just before
the particle arrives at the crossing point. One can imagine either of these
experiments repeated many times, and the result will be that the presence or
absence of $BS_2$ at the crossing point at the instant the particle arrives
there determines whether the particle is always detected in $D^+$ or randomly
detected in $D^+$ and $D^-$. And experimental checks can be carried out with
phase shifters or barriers placed on the paths inside the interferometer. The
paradox is perhaps most telling if one starts off with a series in which $BS_2$
is absent, and the particle arrives randomly in $D^+$ or $D^-$, so about half
the time it is detected in $D^-$, and hence, plausibly, it has been following
the lower path through the interferometer. Now undertake a series of runs in
which $BS_2$ is initially absent, but is inserted in its proper place at the
very last moment. In all of these runs the particle is detected by $D^+$. But
in roughly half of these cases, assuming there is no retrocausal effect from
the later insertion of $BS_2$, the particle must have been traveling through
the lower arm, and were it traveling through the lower arm it would, upon
passing through $BS_2$, arrive with equal probability in either of the
detectors. Thus it might seem that sometimes the particle when traveling
through the lower arm of the interferometer senses that at a future moment
$BS_2$ will be present and decides to split itself into a pair of wavepackets,
one in each arm, with an appropriate phase, so that it will arrive with
certainty at $D^+$. That seems very strange. Is there not some other way of
understanding what is going on without invoking magic or retrocausation?

\begin{figure}[h] 
$$\includegraphics{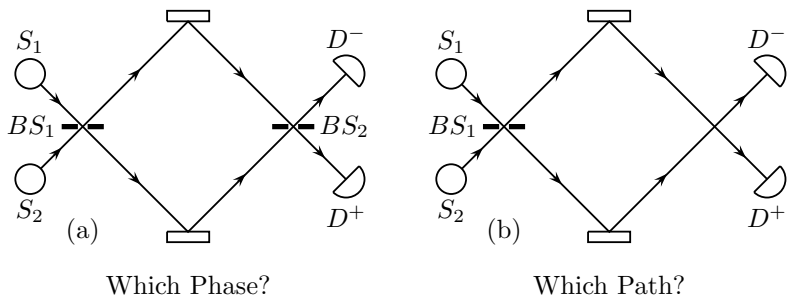}$$
\caption{%
 Mach-Zehnder interferometer with two inputs (a) arranged to determine relative
 phase between the two arms; (b) arranged to measure which path (which arm).
}
\label{fgr4}
\end{figure}

\xb
\outl{MZ with two sources. Which phase vs which path}
\xa

Adding a second source $S_2$ to Wheeler's paradox, Fig.~\ref{fgr4}, makes it
somewhat analogous to our previous discussion of Einstein's paradox. In any
given run, only one source emits a photon, and the phases have been chosen so
that with the second beamsplitter present a particle (photon) which originates
in source $S_1$ will later arrive in $D^+$, and one emitted by $S_2$ will
arrive at $D^-$. In both cases the particle while inside the interferometer is
a superposition of a state $\ket{z+}$ in the upper arm and a state $\ket{z-}$
in the lower arm; in particular let us assume the phases are such that
\begin{equation}
 S_1 \ra (\ket{z+} + \ket{z-})/\st \ra D^+\quad
 S_2 \ra (\ket{z+} - \ket{z-})/\st \ra D^-.
\label{eqn1}
\end{equation}
One can then regard the second beam splitter and the two detectors as forming a
single measurement apparatus that measures ``which phase?''---the difference
between the two possible relative phases, $+$ vs.\ $-$ in \eqref{eqn1}---when
$BS_2$ is in place; or ``which arm?'' if $BS_2$ has been removed. Note the
analogy with the situation depicted in Fig.~\ref{fgr2} (with (a) and (b)
interchanged). The fact that in any particular run the experimenter, by leaving
$BS_2$ in place or removing it can measure which phase or which path, but
cannot determine both, is a fundamental fact of quantum mechanics. Taking it
into account is essential if one is to make progress in resolving the second
measurement problem.

\xb
\subsection{Calibration\label{sbct2.3}}
\xa

\xb
\outl{Calibration of apparatus by sending in particles in known states}
\xa

Competent experimenters check their apparatus in various ways to make sure it
is operating as designed and gives reliable results. There are varieties of
tests, some suggested earlier: placing collimators in various places, removing
beam splitters from a Mach-Zehnder interferometer or placing absorbers in its
arms, etc. If the apparatus is designed to measure the value of some quantity
(observable) $A$ associated with a particle, the simplest form of
\emph{calibration} means preparing many particles with known values of $A$,
thus having the property corresponding to some particular eigenvalue, and
seeing if the measurement outcome (pointer position) corresponds in each case
to the known property. Once the calibration has been carried out the
experimenter can be confident that when a particle of this type is measured by
the apparatus, the outcome will indicate the value of $A$ possessed by the
particle just before it reached the apparatus, even when the particle's prior
history is unknown. Experimenters frequently make assumptions of this kind, and
without it a significant part of experimental physics would be impossible. A
proper quantum mechanical theory of measurement must be able to justify this
practice. In reality things are not always so simple, since apparatus is never
perfect and one may have to account for possible errors; however, for the
present discussion we shall focus on the ideal case in order to get to the
essentials of quantum measurements.


\xb
\section{Properties, Probabilities and Histories}
\label{sct3}
\xa

\xb
\outl{Those familiar with CH can skip this section}
\xa

This section contains a rapid review of material found elsewhere; readers
familiar with consistent histories can skip ahead to Sec.~\ref{sct4}. See \cite{Grff14b} for an
introduction to consistent histories, \cite{Grff02c} for a detailed treatment, and \cite{Grff14}
for extended comments on some conceptual difficulties.

\xb
\subsection{Quantum Properties}
\label{sbct3.1}
\xa

\xb
\xa

\xb
\outl{Physical property: true or false. vN: Property $\lra$ Hilbert
  subspace $\lra$ projector}
\xa

We use the term \emph{physical property} for something like ``the energy is
less than 2 Joules'' or ``the particle is in a region $R$ in space,''
something which can be true or false, and thus distinct from a \emph{physical
  variable} such as the energy or position, represented by a real number in
suitable units. Von Neumann, Sec.~III.5 of \cite{vNmn32b}, proposed that a
quantum property should correspond to a \emph{subspace} of the quantum Hilbert
space, or, equivalently, the \emph{projector} (orthogonal projection operator)
onto this subspace. (We are only concerned here with finite-dimensional Hilbert
spaces for which all subspaces are closed.) What one finds in textbooks is
consistent with von Neumann's prescription, though this is not always clearly
stated.

\xb
\outl{Qm vs Cl properties. Projectors \& indicators }
\xa

A projector, a Hermitian operator equal to its square, is the quantum analog of
an \emph{indicator function} $P(\gm)$ on a classical phase space $\Gm$, a
function that takes the value 1 if at the point $\gm$ the corresponding
physical property is true, or 0 if it is false. For example, the property that
the energy of a harmonic oscillator is less than 2 Joules corresponds to an
indicator $P(\gm)$ equal to 1 for $\gm$ inside, and 0 for $\gm$ outside, an
ellipse centered at the origin of the $(x,p)$ phase plane. A quantum
projector's eigenvalues are $1$ or $0$, which supports the analogy with a
classical indicator. One can make a plausible case that any ``classical''
property of a macroscopic physical object, when viewed in quantum terms, is
represented by a quantum projector on a very high-dimensional subspace of an
enormous Hilbert space. 

\xb
\outl{Projector $[\psi]$ on 1d subspace, $[\psi^0]+[\psi^1]$ for 2d subspace, 
etc.}
\xa

The smallest nontrivial quantum subspace is one-dimensional, consisting of all
complex multiples of a normalized ket $\ket{\psi}$, and the projector is given
by the corresponding Dirac dyad
\begin{equation}
[\psi] = \dya{\psi}.
\label{eqn2}
\end{equation}
We will often make use of this convenient square bracket notation. A projector
on a two-dimensional subspace can be written in the form $[\psi^0] + [\psi^1]$,
where $\ket{\psi^0}$ and $\ket{\psi^1}$ form an orthonormal basis for the
subspace, and similarly for larger subspaces.

\xb
\outl{Negation;  AND for commuting projectors}
\xa

The analogy between quantum projectors and classical indicators also works for
\emph{negation}. The projector corresponding to the property `\NOT\ $P$' is
$I-P$, where $I$ is the identity operator, and the same holds for a classical
indicator when $I$ is understood as the function taking the value 1 everywhere
on the phase space. Given two indicator functions representing properties $P$
and $Q$, their product, which is obviously the same written in either order,
$P(\gm)Q(\gm) = Q(\gm)P(\gm)$, is the indicator for the property $P\ \AND\ Q$.
(Think of ``energy less than one Joule'' \AND\ ``momentum is positive''). But
in the quantum world the product of two projectors $P$ and $Q$ is itself a
projector \emph{if and only if} they commute: $PQ = QP$, and in this case the
product can be associated with the property $P\ \AND\ Q$.

\xb
\outl{Noncommuting projectors. Spin half. Single framework rule}
\xa

But suppose that $P$ and $Q$ do \emph{not} commute, what then? 
Consider a specific example, that of a spin-half particle, where the Hilbert
space is two-dimensional, and spanned by two orthonormal kets $\ket{z^+}$ and
$\ket{z^-}$, eigenvectors of $S_z$, the $z$ component of spin angular
momentum, with eigenvalues $+1/2$ and $-1/2$ in units of $\hbar$. The
projectors 
\begin{equation}
 P^+ = [z^+],\quad P^- = [z^-],
\label{eqn3}
\end{equation}
in the notation used in \eqref{eqn2},
represent these two physical properties; they commute and their product is 0.
Similarly, 
\begin{equation}
 \ket{x^+} = \bigl(\,\ket{z^+} + \ket{z^-}\,\bigr)/\st,\quad
 \ket{x^-} = \bigl(\,\ket{z^+} - \ket{z^-}\,\bigr)/\st,
\label{eqn4}
\end{equation}
are eigenvectors corresponding to the eigenvalues  $+1/2$ and $-1/2$ of the $x$
component of spin angular momentum $S_x$. The corresponding projectors
\begin{equation}
 Q^+=[x^+],\quad Q^-=[x^-]
\label{eqn5}
\end{equation}
commute, and their product is zero. However, neither $Q^+$ nor $Q^-$ commutes
with either $P^+$ or $P^-$. Because the projectors do not commute there is, in
the consistent histories approach, no way to make sense of a statement like ``$S_z=+1/2$ \AND\
$S_x = -1/2$.'' And there is no nontrivial subspace of the Hilbert space which
can be associated with such a combination. (In quantum logic
\cite{BrvN36,Mttl09} one would associate the trivial subspace containing only
the 0 ket with such a conjunction, but quantum logic has its own set of
conceptual difficulties; see \cite{Grff14}.) This is an instance of the
\emph{single framework rule} discussed in more detail in Sec.~\ref{sbct3.3}.

From time to time the claim has been made that the consistent histories approach is logically
inconsistent. However, none of these claims when scrutinized has turned out to
be correct. What typically happens is that the author has either overlooked the
single framework rule or has not taken it seriously. Arguments that show that
consistent histories is internally consistent will be found in Ch.~16 of \cite{Grff02c}, Sec.~4.1
of \cite{Grff14}, and Sec.~8.1
of \cite{Grff13}.

\xb
\subsection{Quantum Probabilities}
\label{sbct3.2}
\xa

\xb
\outl{Sample space $\lra$ Qm projective decomp of identity (PDI); event algebra}
\xa

Ordinary (Kolmogorov) probability theory employs a \emph{sample space} of
\emph{mutually exclusive} items or situations which together
\emph{exhaust all possibilities}, and an \emph{event algebra} which in simple
situations consists of all subsets (including the empty set) of items from the
sample space. In classical statistical mechanics the sample space can consist
of all the distinct points $\gm$ that make up the phase space $\Gm$, but one
could also cut up the phase space into nonoverlapping regions, ``cells'', and
use these for the sample space. The quantum analog of a sample space is a
\emph{projective decomposition of the identity} (PDI): a collection of
projectors $\{P^j\}$ (the superscripts are labels, not exponents) satisfying
\begin{equation}
  I = \sum_j P^j,\quad P^j = (P^j)\ad,\quad P^j P^k = \dl_{jk}P^j.
\label{eqn6}
\end{equation}
Obviously, each projector commutes with every other projector
in the PDI. The simplest choice for a corresponding event algebra, one which
will suffice for our purposes, consists of the 0 projector, all projectors
belonging to the PDI, and in addition all sums of two or more distinct
projectors from the PDI.

\xb
\outl{Physical variable: $A=A\ad$; spectral representation of $A$}
\xa

Given a physical variable $A$ represented by a Hermitian operator $A$ (there is
no harm in using the same symbol for both) there is an associated PDI employed
for the spectral decomposition of $A$,
\begin{equation}
 A = \sum_j \al_j P^j,
\label{eqn7}
\end{equation}
where the eigenvalues $\al_j$ are the possible values which $A$ can take on,
and $P^j$ identifies the subspace where $A$ takes on the value $\al_j$. (We
assume that $\al_j\neq\al_k$ if $j\neq k$ in \eqref{eqn7}; thus for degenerate
eigenvalues the corresponding $P^j$ may project onto a space of dimension
greater than 1.)

\xb
\outl{QM (unlike CM): Choice of sample space = framework is nontrivial }
\xa

In classical physics it is usually the case that only a single sample space
need be considered when discussing a particular physical problem, and so its
choice needs no emphasis, and it may not even be mentioned. In quantum physics
this is no longer the case: many mistakes and numerous paradoxes, e.g., the
Kochen-Specker Paradox (see Sec.~\ref{sbct5.5}), are based on not paying
sufficient attention to the sample space in circumstances in which several
distinct and incompatible sample spaces may seem like reasonable choices. For
this reason it is convenient to use a special term, \emph{framework}, to
indicate the sample space or the corresponding event algebra which is under
discussion.

\xb
\outl{Single framework rule}
\xa

A central feature of consistent histories is the \emph{single framework rule}, which states that
probabilistic reasoning in the quantum context must always be carried out using
a specific and well-defined framework. This rule does \emph{not} prevent the
physicist from \emph{using} many different frameworks when analyzing a
particular physical problem; instead it prohibits \emph{combining} results from
\emph{incompatible} frameworks. Two PDI's $\{P^j\}$ and $\{Q^k\}$ and the
corresponding event algebras are \emph{compatible} provided all the
projectors in one commute with all the projectors in the other: $P^jQ^k = Q^k
P^j$ for every $j$ and $k$. In this case there is a \emph{common refinement}, a
PDI consisting of all nonzero products of the form $P^jQ^k$. Otherwise the
frameworks are incompatible, and the single framework rule prohibits combining
a (probabilistic) inference made using one framework with another that employs
a different framework. If the two frameworks are compatible, then inferences in
one can be combined with those in the other using the common refinement, which
contains both of the event algebras, so again only a single framework is
required. (An additional requirement---consistency conditions---for combining
frameworks arises in the case of quantum histories, Sec.~\ref{sbct3.3}.)

\xb
\outl{Probabilities assigned to PDI. Pure state as pre-probability}
\xa

A PDI can be assigned a probability distribution $p_j = \Pr(P^j)$, where the
$p_j$ are nonnegative real numbers that sum to 1, and this distribution will
generate the probabilities for all the elements in the corresponding event
algebra, just as in ordinary probability theory; e.g., the property $P^1+P^3$ is
assigned the probability $p_1 + p_3$. In quantum mechanics there are various
schemes for assigning probabilities. One method starts with a wavefunction or
pure quantum state $\ket{\psi}$, and assigns to the elements of a PDI $\{P^j\}$
probabilities
\begin{equation}
 p_j = \mte{\psi}{P^j} = \Tr( \,[\psi] P^j ).
\label{eqn8}
\end{equation}
In this situation it is helpful to refer to $\ket{\psi}$ as a
\emph{pre-probability}, i.e., it is used to construct a probability
distribution. Since probability distributions are generally not considered part
of physical reality, at least not in the same sense as physical properties, a
ket or wavefunction used in this way need not be interpreted as something
physical; instead it is simply a tool used to compute probabilities. But in
some other context $\ket{\psi}$ may be a way of referring to the property
represented by the projector $[\psi]$. Carelessly combining these two usages
can cause a great deal of confusion. Note in particular that as long as two of
the $p_j$ in \eqref{eqn8} are nonzero, the property $[\psi]$, or to be more
precise the minimal PDI $\{[\psi], I-[\psi]\}$ that contains it, is
incompatible with the PDI $\{P^j\}$. Hence the single framework rule prevents
using $\ket{\psi}$ as a pre-probability, as in \eqref{eqn8}, while at the same
time regarding it as a physical property of the quantum system.

\xb
\outl{What is the RIGHT framework? Unicity does not hold in Qm world}
\xa

Since the consistent histories interpretation of quantum theory allows many distinct but
incompatible frameworks, a natural question is: Which is the \emph{right}
framework to use in describing some situation of physical interest? In thinking
about this it is helpful to remember that a fundamental difference between
classical and quantum mechanics is that the former employs a phase space and
the latter a Hilbert space for describing a physical system. At a single time a
single point in the phase space represents the ``actual'' state of a classical
system: all properties (subsets of points in the phase space) which contain
this point are \emph{true} and all which do not contain the point are
\emph{false}. The term \emph{unicity} has been used in Sec.~27.3 of
\cite{Grff02c}, and in \cite{Grff14,Grff14b} to describe this concept of a
single unique state of affairs at any given time. However, in the quantum
Hilbert space the closest analogy to a single point in classical phase space is
a one-dimensional subspace or \emph{ray}. If one assumes that one particular
ray is true, then one might suppose that all rays orthogonal to it are false.
But there are many rays that are neither identical to nor orthogonal to the ray
in question; what shall be said of them? Thus attempting to extend the concept
of unicity into the quantum domain runs into problems. We have good reason to
believe that physical reality is better described by quantum theory than by
classical physics, and hence certain classical concepts must be abandoned, to
join others, such as the earth immobile at the center of the universe, which
modern science has rendered untenable, even though for certain purposes they
may remain useful approximations. Unicity seems to belong to that category.

\xb
\outl{Different ``aspects'' of Qm world require appropriate frameworks }
\xa

But the question remains: what are the criteria which lead to the use of a
particular framework, rather than another which is incompatible with it? The
examples in Sec.~\ref{sct2} and various applications in Sec.~\ref{sct5} suggest
that quantum physical situations possess what one might call \emph{different
  aspects}, and a quantum description of a particular aspect can only be
constructed using a framework compatible with that aspect. For example, the
$S_z$ ``aspect'' of a spin half particle can only be discussed using the $S_z$
framework; the $S_x$ framework is of no use. As is usual with with unfamiliar
concepts, the best way to understand them is to apply them to several different
examples. In particular, in Secs.~\ref{sbct5.1} and \ref{sbct5.2} we will show
how the use of frameworks can ``untangle'' the paradoxes in Secs.~\ref{sbct2.1}
and \ref{sbct2.2}.

\xb
\subsection{Histories and the Extended Born Rule}
\label{sbct3.3}
\xa

\xb
\outl{History = sequence of properties. Family of histories = PDI of $\breve I$}
\xa

A quantum history is best understood as a \emph{sequence of quantum properties
  at successive times}. A classical analogy is a sequence of coin tosses, or
rolls of dice. The theory is simplest if one employs a finite set of discrete
times, rather than continuous time. This is no real limitation, as these
times may be arbitrarily close together. A history associated with the times
$t_0 < t_1 < t_2 <\cdots t_n$ can be written in the form
\begin{equation}
 Y = F_0 \od F_1 \od F_2 \od \cdots F_n,
\label{eqn9}
\end{equation}
where each $F_j$ is a projector representing some quantum property at the time
$t_j$, and the $\od$ separating properties at successive times
are tensor product symbols, a variant of $\ot$. Thus if $\HC$ is the quantum
Hilbert space at one time, $Y$ in \eqref{eqn9} is a projector on the tensor
product \emph{history} Hilbert space $\breve \HC = \HC^{\ot (n+1)}$. A
\emph{family} of histories is a collection of such projectors that sum to the
history identity $\breve I = I\od I\od \cdots I$, thus a PDI.
For present purposes it suffices to use a family in which the histories
are of the form
\begin{equation}
 Y^\al = [\Psi_0] \od F_1^{\al_1} \od F_2^{\al_2} \od \cdots F_N^{\al_n}.
\label{eqn10}
\end{equation}
where $[\Psi_0]$, see \eqref{eqn2}, is the projector on a pure state
$\ket{\Psi_0}$. The superscripts are labels distinguishing different projectors
at the same time, and together they form a vector $\al = (\al_1, \al_2,\ldots
\al_n)$. In addition there is a special history $Y^0 = I-[\Psi_0] \od I \od I
\cdots I$ which is assigned zero probability, and whose sole purpose is to
ensure that the history projectors sum to $\breve I$.

\xb
\outl{Complete family defined. We won't need projectors entangled between 
different times}
\xa

A \emph{complete} family of histories is one in which the $Y^\al$ sum 
to $\breve I$, but we will also use the term if they sum to $\breve I - Y^0$.
One way to ensure that the family is complete is if for each time $t_j > t_0$
it is the case that the $\{F_j^{\al_j}\}$ are a PDI of $\HC$, but this is often
too restrictive. There is no reason why a family should not contain projectors
on states ``entangled'' between different times, but in the following
discussion we will only need ``product'' histories as in \eqref{eqn9}.

\xb
\outl{Assigning probabilities. Chain kets for closed system. Consistency}
\xa

Since a family of histories is a PDI it can serve as a probabilistic sample
space for the quantum analog of a classical stochastic process such as a random
walk. As in the classical case there is no fixed rule for assigning
probabilities to such a process. However, in a \emph{closed} quantum system for
which Schr\"odinger's equation yields unitary time development operators
$T(t',t)$ (e.g., $\exp[ -i(t'-t)H/\hbar]$ in the case of a time-independent
Hamiltonian $H$) these can be used to assign probabilities to a history family
using an extension of the Born rule, provided certain \emph{consistency}
(or \emph{decoherence}) \emph{conditions} are satisfied. If all histories start
with the same initial pure state one defines a \emph{chain ket} (an element of
$\HC$ not $\breve \HC$):
\begin{equation}
 \ket{Y^\al} = F_n^{\al_n} T(t_n,t_{n-1}) F_{n-1}^{\al_{n-1}}T(t_{n-1},t_{n-2})
 \cdots  F_1^{\al_1} T(t_1,t_0) \ket{\Psi_0}.
\label{eqn11}
\end{equation}
The consistency conditions are the requirement that the chain kets are
orthogonal for distinct histories, 
\begin{equation}
 \inpd{Y^\al}{Y^{\al'}} = 0 \text{ for } \al\neq\al',
\label{eqn12}
\end{equation}
When it is satisfied the extended Born rule assigns to each history of the
sample space a probability
\begin{equation}
 \Pr(Y^\al) = \inp{Y^\al}.
\label{eqn13}
\end{equation}

\xb
\outl{Orthogonal chain kets made plausible. 2-time family always consistent}
\xa

The orthogonality requirement \eqref{eqn12} is not unnatural when one remembers
that the $\ket{Y^\al}$ are elements of the single-time Hilbert space $\HC$, not
the history space $\breve \HC$, and the ordinary Born rule is used to assign
probabilities to an orthonormal basis, or, more generally, a PDI. In fact, for
a history involving only two times, $t_0$ and $t_1$, the consistency condition
is automatically satisfied because the $F_1^{\al_1}$ for different $\al_1$ form
a PDI on $\HC$, and then \eqref{eqn13} is just the usual Born probability.

\xb
\outl{QM allows DESCRIPTION of INDIVIDUAL processes, not just averages}
\xa

It is important to notice that quantum mechanics allows a \emph{description} of
what happens in an individual realization of a quantum stochastic process, even
though the dynamics is probabilistic; the same as in a classical stochastic
theory. One is sometimes given the impression that quantum theory only allows a
discussion of statistical averages over many runs of an experiment. This is not
the case, and it is easy to identify instances where individual outcomes and
not just averages play a significant role. For example, if Shor's quantum
algorithm \cite{Shr94,Grjy05} is employed to factor a long integer, then at the
end of each run the outcome of a measurement is processed to see if this result
solves the problem, and if it does, no further runs are needed. While it may
take more than one run to achieve success, the outcome of a particular run is a
significant quantity, and not just the average over several runs. Similarly, in
the case of Einstein's paradox, Sec.~\ref{sbct2.1}, a flash of light at a
particular point on the fluorescent screen, Fig.~\ref{fgr1}(a), can be
understood to mean that the particle traveled on a straight (or almost
straight) path from the source to the screen on this particular occasion.


\xb
\section{Measurement Models \label{sct4}}
\xa

\xb
\subsection{Projective Measurements \label{sbct4.1}}
\xa

\xb
\outl{Modified vN model. Time development $T(t',t)$. Initial $\ket{\psi_0}\ot
  \ket{\Phi_0}$. Density operators vs. pure states }
\xa

Our first model is a generalization of the one introduced by von Neumann in
Sec.~VI.3 of \cite{vNmn32b}.%
\footnote{%
  Von Neumann also gives a specific application of his general model to the
  case of a ``Gaussian probe'' whose momentum is shifted by an (almost)
  instantaneous interaction with the measured system. Our discussion concerns
  the more general model rather than its application to the Gaussian probe.} %
Let $\HC_s$ be the Hilbert space of the system to be measured, which for
convenience will hereafter be referred to as ``the particle'', whereas the
measuring apparatus, including its environment if that is significant, is
described by a Hilbert space $\HC_m$. The total system with Hilbert space
$\HC_M =\HC_s\ot \HC_m$ is thought of as closed, so its dynamics can be
associated with a collection of unitary time development operators $T(t',t)$.
We will focus on histories involving three times $t_0 < t_1 <t_2$, where the
interval from $t_0$ to $t_1$ is so short that $T(t_1,t_0)\approx I$ and thus
\begin{equation}
 T(t_2,t_0) \approx T(t_2,t_1)
\label{eqn14}
\end{equation}
with negligible error. At the initial time $t_0$ the particle can be assigned a
quantum state $\ket{\psi_0}$ in $\HC_s$, and the apparatus (and environment) a
state $\ket{\Om_0}$ in $\HC_m$; hence an initial state
\begin{equation}
 \ket{\Psi_0} = \ket{\psi_0}\ot \ket{\Om_0}.
\label{eqn15}
\end{equation}
for the combined, closed system. 
The use of pure states rather than density operators does not involve any loss
of generality; see Sec.~\ref{sbct4.4} for additional comments. But the
requirement that $\ket{\Psi_0}$ in \eqref{eqn15} be a \emph{product state} is
important. It means that the particle and the apparatus (or environment) are
initially uncorrelated, at least to a sufficiently good approximation.

\xb
\outl{Particle \& apparatus interact during $[t_1,t_2]$. PDI $\{M^k\}$ for
  pointer positions}
\xa

We assume that the interaction between the particle and the apparatus
takes place during the time interval between $t_1$ and $t_2$, and as a
consequence of this interaction 
\begin{equation}
 T(t_2,t_1) \bigl(\ket{s^j} \ot \ket{\Om_0}\bigr) = 
\ket{\Phi^j},
\label{eqn16}
\end{equation}
where the $\ket{s^j}$ form an orthonormal basis for the particle Hilbert space
$\HC_s$, while the $\ket{\Phi^j}$, which lie in the Hilbert space $\HC_M$, are
states of the particle plus apparatus that correspond to distinct macroscopic
outcomes of the measurement---distinct ``pointer positions'' of the apparatus,
to use the traditional terminology of quantum foundations---in the sense of
satisfying \eqref{eqn17} below. (The space $\HC_M$ has the same dimension as
$\HC_s\ot\HC_m$, but we have not written it in that form since sometimes the
particle does not even exist at the end of the measurement. See the discussion
of nondestructive measurements in Sec.~\ref{sbct4.3}.) These pointer positions
are mutually orthogonal, as is always the case for states which are
macroscopically distinct. To be more precise, we assume there is a PDI
$\{M^k\}$ on $\HC_M$ such that
\begin{equation}
 M^k \ket{\Phi^j} = \dl_{jk} \ket{\Phi^j},
\label{eqn17}
\end{equation}
where each $M^k$ is a projector on a macroscopic subspace (property) whose
interpretation is that the pointer is in position $k$, and \eqref{eqn17} says
that $\ket{\Phi^k}$ lies within the subspace defined by $M^k$. To ensure that
the $\{M^k\}$ sum to the identity on $\HC_M$, assume that the possible
pointer positions are represented by $k=1,2,\ldots n$, and let
\begin{equation}
 M^0 := I_M - \sum_{k=1}^n M^k
\label{eqn18}
\end{equation}
project on the subspace that includes all other possibilities (e.g., the
apparatus has broken down).

\xb
\outl{Isometry $J:\HC_s \ra \HC_M$. Backwards map $[s^k] = J\ad M^k J$}
\xa

To better understand what this measurement measures it is useful to
introduce an \emph{isometry} $J:\HC_s \ra \HC_M$ defined by
\begin{equation}
 J\ket{\psi} = T(t_2,t_1) \bigl(\ket{\psi}\ot\ket{\Om_0}\bigr).
\label{eqn19}
\end{equation}
An isometry, like a unitary, preserves lengths, and is
characterized by the requirement that
\begin{equation}
 J\ad J = I_s,
\label{eqn20}
\end{equation}
where $J\ad: \HC_M \ra \HC_s$ is the adjoint of $J$. (The operator $J J\ad:
\HC_M\ra\HC_M$ is a projector on the subspace of $\HC_M$ that is the image of
under $J$ of $\HC_s$, and is not important for our discussion.) 

The isometry that corresponds to $T(t_2,t_1)$ in \eqref{eqn16} is
\begin{equation}
J\ket{s^j} = \ket{\Phi^j}.
\label{eqn21}
\end{equation}
Combining this with \eqref{eqn17} leads to
\begin{equation}
 M^k J\ket{s^j} = \dl_{jk} J\ket{s^j}.
\label{eqn22}
\end{equation}
Multiplying both sides on the left by $J\ad$ and using \eqref{eqn20} yields
\begin{equation}
 J\ad M^k J\ket{s^j} = \dl_{jk}\ket{s^j},
\label{eqn23}
\end{equation}
which implies that 
\begin{equation}
 [s^k] = \dya{s^k} = J\ad M^k J.
\label{eqn24}
\end{equation}
That is, the ``backwards map'' $J\ad( \cdot) J$ applied to the projector $M^k$
on the subspace that corresponds to pointer position $k$ is the prior
microscopic state $[s^k]$ giving rise to this outcome.

\xb
\outl{Family $Y^k = [\psi_0]\ot[\Om_0] \od I \od M^k$; $\Pr(M_2^k) = \inp{Y^k}
  = |c_k|^2$ }
\xa

To complete the discussion of projective measurements we need to introduce 
families of histories. Let us begin with the family $\{Y^k\}$ consisting of 
histories
\begin{equation}
 Y^k = [\Psi_0] \od I \od M^k 
\label{eqn25}
\end{equation}
at times $t_0<t_1<t_2$, where $\ket{\Psi_0}$ was defined in \eqref{eqn15}, and
\begin{equation}
 \ket{\psi_0} =\sum_j c_j \ket{s^j},\quad 
\label{eqn26}
\end{equation}
is an arbitrary state of $\HC_s$. 
The chain kets 
\begin{equation}
 \ket{Y^k} = c_k\ket{\Phi^k}
\label{eqn27}
\end{equation}
associated with these histories (remember that $T(t_2,t_1) \approx T(t_2,t_0)$)
are obviously orthogonal to each other in view of \eqref{eqn17} and the fact
that the $\{M^k\}$ form a PDI. Thus the Born rule assigns a probability
\begin{equation}
 \Pr(M_2^k) = \inp{Y^k} = |c_k|^2,
\label{eqn28}
\end{equation}
the absolute square of the coefficient of $\ket{\psi_0}$ in \eqref{eqn26}, to
the pointer outcome $k$, in agreement with textbooks, but without employing any
special rule for measurements, since \eqref{eqn28} is nothing but a particular
application of the general formula \eqref{eqn13} that assigns probabilities to
histories.

\xb
\outl{First measurement problem (Schr cat) is not present}
\xa

Note that the \emph{first} measurement problem, attempting to give a physical
interpretation to the macroscopic superposition state
\begin{equation}
 \ket{\Psi_2} = T(t_2,t_0) \ket{\Psi_0} = \sum_j c_j\ket{\Phi^j},
\label{eqn29}
\end{equation}
never arises, because $\ket{\Psi_2}$ has never entered the discussion. To be
sure, from the consistent histories perspective there is nothing wrong with the family consisting
of just the two histories
\begin{equation} [\Psi_0] \od I \od \{[\Psi_2],I-[\Psi_2]\},
\label{eqn30}
\end{equation} 
where each history uses one of the projectors inside the curly brackets. 
It (trivially) satisfies the consistency condition, and the Born rule
assigns a probability of 1 to $[\Psi_2]$. It is a perfectly good quantum
description which, however, is incompatible with the family \eqref{eqn25} if at
least two of the $c_j$ in \eqref{eqn26} are nonzero, since $[\Psi_2]$ will then
not commute with the corresponding $M^j$, rendering a discussion of measurement
outcomes impossible. Combining the families in \eqref{eqn25} and \eqref{eqn30}
is as silly as simultaneously assigning to a spin-half particle a value for
$S_z$ along with one for $S_x$. The choice of which of these families to use
will generally be made on pragmatic grounds. In particular, if one wants to
discuss real experiments of the sort actually carried out in laboratories and
what one can infer from their outcomes---one might call this practical
physics---the choice is clear: one needs to employ a family in which
measurements have outcomes. 

\xb
\outl{Response to objections to framework choice}
\xa

There are physicists who object to a framework choice based on pragmatic
grounds which seem related to human choice, e.g., see Sec.~3.7 of
\cite{Wnbr15}, though they might not object to astronomers interested in the
properties of Jupiter using concepts appropriate to that planet rather than,
say, Mars. Of course this is a classical analogy, but thinking about it, along
with the spin-half example mentioned earlier, may help in understanding how the
single framework rule can assist in sorting out quantum paradoxes while still
allowing quantum theory to be an objective science. The idea that there can
only be exactly one valid quantum description, the principle of unicity
discussed in Sec.~\ref{sbct3.2}, runs into difficulties in the case of
Einstein's paradox, Sec.~\ref{sbct2.1}, as well rendering the infamous first
measurement problem insoluble for reasons that have just been discussed.

\xb
\outl{Microscopic $[s^j]$ at intermediate $t_1$: family 
$\{Y^{jk}\}$ $\ra$ $\Pr(s^j_1 \vbl M^k_2) = \dl_{jk}$}
\xa

After this diversion let us return to the second measurement problem.
To see how the macroscopic measurement outcomes $M^k$ are related to the 
microscopic properties the measurement was designed to measure,
we introduce a refinement $\{Y^{jk}\}$ 
\begin{equation}
 Y^{jk} = [\Psi_0] \od [s^j] \od M^k,
\label{eqn31}
\end{equation}
of the family \eqref{eqn25} considered previously. Here $[s^j]$ at the
intermediate time $t_1$ is to be interpreted, following the usual physicists'
convention, as $[s^j]\ot I_m$; the property $[s^j]$ of the particle and no
information about anything else. The corresponding chain kets, see
\eqref{eqn26} and \eqref{eqn16},
\begin{equation}
 \ket{Y^{jk}} = c_j \dl_{jk}\ket{\Phi^k}.
\label{eqn32}
\end{equation}
are mutually orthogonal since the $\ket{\Phi^k}$ are orthogonal. Thus
the family $\{Y^{jk}\}$ is consistent, and yields a joint probability
distribution
\begin{equation}
 \Pr(s^j_1 ,M^k_2) = \inp{Y^{jk}} = \dl_{jk} |c_j|^2,
\label{eqn33}
\end{equation}
where the subscripts of the arguments of $\Pr()$ indicate time. Summing over
$j$ gives \eqref{eqn28}, and combining that with \eqref{eqn33} yields
conditional probabilities:
\begin{equation}
 \Pr(s^j_1 \vbl M^k_2) =  \Pr(s^j_1,M^k_2)/\Pr(M^k_2) = \dl_{jk},
\label{eqn34}
\end{equation}
assuming $c_k$ is nonzero. In words: if the measurement outcome (pointer
position) is $k$, i.e., $M^k$, at time $t_2$, the particle certainly had the
property $[s^k]$ at time $t_1$. Thus the second measurement problem has been
solved for the case of projective measurements. Note that this conclusion does
\emph{not} depend upon the initial state $\ket{\psi_0}$, which only determines
the probability of the measurement outcome $M^k$ as noted above in
\eqref{eqn28}. (If $c_k=0$, \eqref{eqn34} does not hold, but it is also not
needed, since the outcome $k$ will never occur.)

\xb
\subsection{Generalized Measurements and POVMs \label{sbct4.2}}
\xa

\xb
\outl{Setup: $J$ and $\{M^k\}$ as before; define $Q^k := J\ad M^k J$}
\xa

The basic setup for discussing generalized measurements is the same as that in
Sec.~\ref{sbct4.1}: times $t_0 < t_1 < t_2$, an initial state \eqref{eqn15} at
time $t_0$, negligible time development (see \eqref{eqn14}) between $t_0$ and
$t_1$, the isometry $J$ defined in \eqref{eqn19}, and a PDI $\{M^k\}$
corresponding to different pointer positions at $t_2$. However, we now drop the
assumption of an orthonormal basis $\{\ket{s^j}\}$ of $\HC_s$ with $J\ket{s^j}$
lying in the space $M^j$. Instead, use the backwards map of the projectors on
the pointer subspaces to \emph{define} for each $k$ an operator
\begin{equation}
 Q^k := J\ad M^k J
\label{eqn35}
\end{equation}
on $\HC_s$. For a projective measurement $Q^k= [s^k]$ is the property possessed
by the particle at the earlier time $t_1$ when the measurement outcome is
$M^k$, and we shall see that something similar, though a bit more complicated,
holds for generalized measurements. Another special case, a \emph{generalized
  projective measurement}, is one in which each $Q^k$ is a projector and
the $\{Q^k\}$ form a PDI, but one or more may have a rank (so project on a
subspace of dimension) greater than 1.

\xb
\outl{Proof that $\{Q^k\}$ is a POVM}
\xa

The collection $\{Q^k\}$ forms a \emph{POVM} (positive operator-valued
measure), a collection of positive semi-definite operators with sum equal to
the identity on $\HC_s$. The equality
\begin{equation}
 \mte{\psi}{Q^k} = \mte{\psi}{J\ad M^k J} = \mte{\Psi}{M^k} \geq 0,
\label{eqn36}
\end{equation}
for an arbitrary $\ket{\psi}$ in $\HC_s$, with $\ket{\Psi}=J\ket{\psi}$,
demonstrates that $Q^k$, just like the projector $M^k$, is a positive
semi-definite operator. Summing both sides of \eqref{eqn36} over $k$ and
remembering that the $M^k$ form a PDI shows that
\begin{equation}
 \sum_k Q^k = I_s,
\label{eqn37}
\end{equation}
completing the proof that $\{Q^k\}$ is a POVM. (Note that the special $M^0$ in
\eqref{eqn18} gives rise to $Q^0=0$.)

\xb
\outl{2d measurement problem for POVM: Use spectral rep 
$Q^k = \sum_j q_{jk}  \xi^{jk}$ }
\xa

The first measurement problem for such a POVM is solved in exactly the same way
as for the von Neumann model: use the PDI $\{M^k\}$ at time $t_2$, not
the projector $[\Phi_2]$ of the unitarily evolved state. The second measurement
problem is more subtle, as it requires introducing suitable properties as
events at $t_1$ to produce a consistent family. The choice is not unique, but
the following is a quite general and fairly useful approach. The spectral
decomposition of $Q^k$ can be written in the form
\begin{equation}
Q^k = \sum_j q_{jk} \xi^{jk};\quad \sum_j \xi^{jk} = I_s,
\label{eqn38}
\end{equation}
where for each fixed $k$ the $\xi^{jk}$ labeled by $j$ are projectors that form
a PDI on $\HC_s$, while the $q_{jk}\geq 0$ are the corresponding eigenvalues of
$Q^k$.  We assume the eigenvalues are unique, $q_{jk} \neq
q_{j'k}$ when $j\neq j'$, so some of the $\xi^{jk}$ may have rank greater than
one. As with any PDI the projectors are orthogonal
and sum to the identity:
\begin{equation}
 \xi^{jk} \xi^{j'k} = \dl_{jj'} \xi^{jk},\quad
 \sum_j \xi^{jk} = I_s. 
\label{eqn39}
\end{equation}

\xb
\outl{Histories family $Y^{jk} = [\Psi_0] \od \xi^{jk} \od M^k$ is consistent}
\xa

The family $\{Y^{jk}\}$ of histories 
\begin{equation}
 Y^{jk} = [\Psi_0] \od \xi^{jk} \od M^k
\label{eqn40}
\end{equation} 
when augmented with the uninteresting $[\Psi_0] \od I \od M^0$ (of zero weight),
is complete, since 
\begin{equation}
\sum_j Y^{jk} =[\Psi_0] \od I \od M^k.
\label{eqn41}
\end{equation}
The chain kets
\begin{equation}
 \ket{Y^{jk}} = M^k J\, \xi^{jk}\, \ket{\psi_0}
\label{eqn42}
\end{equation}
are obviously mutually orthogonal if the two $k$ values differ. For a given $k$
we need to consider
\begin{equation}
 \inpd{Y^{jk}}{Y^{j'k}} = \mte{\psi_0}{\xi^{jk} J\ad M^k J \xi^{j'k}}
 =\mte{\psi_0}{\xi^{jk} Q^k \xi^{j'k}} = 
 \dl_{jj'} q_{jk}\mte{\psi_0}{\xi^{jk}},
\label{eqn43}
\end{equation}
where the second equality follows from \eqref{eqn35} the third from
\eqref{eqn38} and \eqref{eqn39}.
Thus the family $\{Y^{jk}\}$ defined in \eqref{eqn40} is consistent, with
probabilities
\begin{equation}
  \Pr(\,\xi_1^{jk}, M_2^{k'}) = \dl_{kk'}\, q_{jk}\, \mte{\psi_0}{\xi^{jk}},
\label{eqn44}
\end{equation}
where subscripts 1 and 2 identify the times $t_1$ and $t_2$ before and
after the measurement takes place. It follows that
\begin{align}
 \Pr(M^k_2) &= \sum_j \Pr(\,\xi_1^{jk}, M_2^k) = \mte{\psi_0}{Q^k},
\label{eqn45}\\
\Pr(\,\xi_1^{jk'}\vbl M_2^{k}) &= 
 \dl_{kk'}\, q_{jk}\, \mte{\psi_0}{\xi^{jk}}/\mte{\psi_0}{Q^k}. 
\label{eqn46}
\end{align}

\xb
\outl{Inferences from measurement outcomes}
\xa

What \eqref{eqn46} tells us is that if the outcome (pointer position) is $k$
the system earlier had one of the properties $\xi^{jk}$, with probabilities
that will in general depend on the initial particle state $\ket{\psi_0}$. If
$Q^k$ is itself a projector or proportional to a projector, as will be the case
for a general projective measurement, one can be sure that the particle
possessed the property $Q^k$ at time $t_1$. If the support of $Q^k$ is a proper
subspace of $\HC_s$, the system can be assigned the property corresponding to
this subspace at the time immediately before the measurement. If neither of
these conditions holds it may be possible on the basis of additional
information about $\ket{\psi_0}$ to assign probabilities to the different
$\xi^{jk}$ for this $k$, or perhaps argue that some of these probabilities are
negligible, allowing one with reasonable confidence to say something nontrivial
about the property possessed earlier by the particle.

\xb
\outl{Consistent family used here is a natural choice; others are possible}
\xa

Note that whereas for a fixed $k$ the $\xi^{jk}$ for different $j$ are mutually
orthogonal, for \emph{different} $k$ values, different outcomes of the
experiment, one may be able to draw different and perhaps mutually incompatible
conclusions about the prior properties. This is a feature of quantum
measurements which has given rise to a lot of confusion, and is best discussed
in terms of a specific example; see the one in Sec.~\ref{sbct5.3}. While the
consistent family in \eqref{eqn40} is not the only possibility for discussing
what one can learn about the prior state of the particle from measurement
outcomes, it is a rather natural choice, especially when nothing else is known
about the measured system.

\xb
\subsection{Nondestructive Measurements and Preparations \label{sbct4.3}}
\xa

\xb
\outl{Define: preparation, nondestructive measurement }
\xa

\xb
\outl{Von Neumann model: $J \ket{s^j} = \ket{s^j} \ot \ket{\Phi^j}$; 
$M^k\ket{\Phi^j} =  \dl_{jk} \ket{\Phi^j}$ }
\xa

\xb
\outl{Pointer position indicates state of particle before and after measurement}
\xa

A \emph{measurement} determines a past property whereas a \emph{preparation} is
a procedure to prepare a particular quantum state, and a \emph{nondestructive}
measurement combines the two: the apparatus both measures and prepares certain
properties. While preparations lie somewhat outside the scope of the present
paper, it is worthwhile making some remarks on the subject, if only because of
the confusion found in textbooks and other publications, where ``measurement''
is often (incorrectly) defined as something that has to do with ``wavefunction
collapse.'' The confusion goes back to von Neumann's original measurement model
in which, using the notation of the present paper, $\HC_M=\HC_s\ot\HC_m$, and
the isometry $J$ in \eqref{eqn19} takes the form 
\begin{equation}
 J \ket{s^j} = \ket{s^j} \ot \ket{\Phi^j},\quad 
M^k\ket{\Phi^j} = \dl_{jk} \ket{\Phi^j}, 
\label{eqn47}
\end{equation}
with the $\{\ket{s^j}\}$ an orthonormal basis of $\HC_s$.
(The $\ket{\Phi^j}$ and the PDI $\{M^k\}$ now refer to $\HC_m$ rather
than $\HC_M$, as in our earlier discussion,  but this is a minor difference.)
In place of \eqref{eqn31} use the family
\begin{equation}
 Y^{jj'k} = [\psi_0]\ot[\Om_0] \od \{[s^j]\} \od \{[s^{j'}]\ot [M^k]\}.
\label{eqn48}
\end{equation}
It is straightforward to show that it is consistent, since all the chain kets
vanish except for the cases $j=j'=k$, with the result
\begin{align}
 &\Pr(s^j_1,s^{j'}_2,M^k_2) = \dl_{jj'}\dl_{jk}\mte{\psi_0}{\,[s^j]\,},
\label{eqn49}\\
 &\Pr(s^j_1\vbl M^k_2) = \dl_{jk},\quad \Pr(s^j_2\vbl M^k_2) = \dl_{jk}.
\label{eqn50}
\end{align}

\xb
\outl{Measurement nondestructive since $[s^k]$ same at $t_1$ and $t_2$}
\xa

\xb
\outl{Wavefunction collapse, $\ket{\psi_0}$ replaced by $\ket{s^k}$, is not
  needed}
\xa

This measurement is nondestructive in the sense that from the outcome $M^k$ one
can immediately infer that the particle property both before and after the
measurement was $[s^k]$, so it did not change. Furthermore, this conclusion is
independent of the initial particle state $\ket{\psi_0}$ (assuming only that
$c_k$ in \eqref{eqn26} is not zero; if it is zero the outcome $M^k$ will never
occur). That the earlier $\ket{\psi_0}$ is replaced by the later $\ket{s^k}$ in
the case of outcome $M^k$ is the idea of ``wavefunction collapse,'' a
confusing notion best replaced with the second equality in \eqref{eqn50}.

\xb
\outl{Measurement-preparations using Kraus operators $K^j$}
\xa

Discussions of measurements are sometimes based on a generalization of
\eqref{eqn47} in which for any $\ket{\psi}$ in $\HC_s$ the isometry is assumed
to be of the form
\begin{equation}
 J\ket{\psi} = \sum_j K^j\ket{\psi}\ot \ket{\Phi^j},
\label{eqn51}
\end{equation}
where the $\{\Phi^j\}$ are an orthonormal collection, and the \emph{Kraus
  operators} $K^j$ (note that $j$ is a label) are arbitrary maps of $\HC_s$ to
itself subject only to the condition that
\begin{equation}
 \sum_j (K^j)\ad K^j = I_s,
\label{eqn52}
\end{equation}
which guarantees that $J$ in \eqref{eqn51} is an isometry. Regarded as a
measurement, which is to say something that determines the property of the
particle at $t_1$, this is equivalent to a POVM in which 
\begin{equation}
 Q^j = (K^j)\ad K^j.
\label{eqn53}
\end{equation}

\xb
\outl{L\"uders' model: the $K^j$ are projectors}
\xa

The nondestructive model in \eqref{eqn47} is easily extended to a general PDI
$\{P^j\}$ on $\HC_s$ by setting the Kraus operator $K^j$ in \eqref{eqn51} equal
to $P^j$, whence it follows that any initial $\ket{\psi}$ in $\HC_s$ with the
property $P^k$, i.e., $P^k\ket{\psi}=\ket{\psi}$ will result in a measurement
outcome $M^k$ and $\ket{\psi}$ will emerge unchanged at time $t_2$. This is the
essence of L\"uders' proposal \cite{Ldrs06,BsLh09b}, which is best regarded as
a particular model of a nondestructive measurement and not (as sometimes
supposed) a general principle of quantum theory.

\xb
\outl{Preparation without measurement. Prepared states need not be orthogonal}
\xa

\xb
\outl{Selecting the desired state from a stochastic preparation}
\xa

In the case of a preparation one is not interested in the property of the
particle at an earlier time, but instead its state at a time $t_2$ after the
interaction with the measuring device is over. If, for example, the isometry is
given by \eqref{eqn47}, then according to \eqref{eqn50} if the pointer is in
position $k$ at time $t_2$ one can be certain that the particle is in state
$[s^k]$ at this time. But a simpler and more general preparation model is
obtained if in place of \eqref{eqn47} one assumes there is a normalized state
$\ket{\psi_1}$ at time $t_1$ and an isometry $J$ such that
\begin{equation}
 J\ket{\psi_1} = \sum_k \sqrt{p_k}\, \ket{\hat s^k}\ot\ket{\Phi^k},\quad 
M^k\ket{\Phi^{k'}} = \dl_{kk'} \ket{\Phi^k},
\label{eqn54}
\end{equation}
where the $p_k$ are probabilities that sum to 1. The states $\ket{\hat s^k}$
are normalized, but we do not assume that they form a basis; in particular,
they need not be mutually orthogonal. Nonetheless one can infer that if at
$t_2$ the pointer is in position $k$, the particle at this time is in the state
$\ket{\hat s^k}$. Note that even if the $\ket{\hat s^k}$ are not orthogonal the
states $\ket{\hat s^k}\ot\ket{\Phi^k}$ are orthogonal and hence distinct; see
Ch.~14 in \cite{Grff02c} for some discussion of states of this sort. One might
worry that this preparation model is stochastic: if outcome $k=3$ is desired,
sometimes it will occur and sometimes it won't. But since the pointer position
is macroscopic it is not difficult to design a system whereby undesired
outcomes are removed (e.g., run the particle into a barrier), or if one is
repeating the experiment many times, simply keep a record of the value of k for
each run, and throw out the runs for which it is not equal to 3.


\xb
\subsection{Some Remarks About Density Operators \label{sbct4.4}}
\xa

\xb
\outl{Density operator $\rho \lra$ ensemble. Replace $\mte{\psi_0}{W}$
with $\Tr(\rho W)$}
\xa

The foregoing discussion of measurement models employed pure states and
projectors on pure states, and it is natural to ask what the appropriate
formulation ought to be if one is dealing with mixed states. Mixed states arise
in quantum mechanics in two somewhat different ways. The first is analogous to
a classical probability distribution: one has in mind some collection of pure
states $\ket{\psi^j}$ with associated probabilities $p_j$, known as an
\emph{ensemble}, and the associated density operator is
\begin{equation}
 \rho = \sum_j p_j [\psi^j].
\label{eqn55}
\end{equation}
Suppose particles are prepared in states chosen from this ensemble with the
specified probabilities, and then measured. What can one infer about the state
of a particle just before the measurement, given a particular outcome? Since
the only role of the initial state $\ket{\psi_0}$ in Secs.~\ref{sbct4.1} and
\ref{sbct4.2} is to assign probabilities, in the case of a random input one
replaces $\ket{\psi_0}$ by $\rho$ when computing averages; e.g.,
$\mte{\psi_0}{Q^k}$ in \eqref{eqn45} is replaced with $\Tr(\rho\, Q^k)$. Note
that the state inferred in this way from the measurement outcome in a
particular run need not be the same as the member of the ensemble sent into the
measurement apparatus. This is no more surprising than the fact that the
$[s^k]$ inferred in \eqref{eqn38} can be different from $[\psi_0]$.

\xb
\outl{Density operator from partial trace}
\xa

The second way in which a density operator arises is through taking a partial
trace of an entangled pure state on a composite system down to one of the
subsystems; see Ch.~15 of \cite{Grff02c} for further details. If one is only
concerned with properties of this particular subsystem and not its correlations
with the others, and if only this subsystem interacts with the measuring
apparatus, then the previous discussion applies: the situation is exactly the
same as for the case of an ensemble. If, however, one is interested in
correlations with the another subsystem or subsystems it is best to treat the
entire system under consideration as a single system when working out what one
can infer from a measurement, even if the measurement is carried out on just
one of the subsystems, as the density operator may not provide the sort of
information one is interested in. See Sec.~\ref{sbct5.6} below for an example.

\xb
\outl{Replacing initial apparatus $\ket{\Om_0}$ with density operator. 
See Ch.~17 of CQT}
\xa

One may also be concerned about using a pure initial state $\ket{\Om_0}$ for a
macroscopic apparatus rather than a density operator or a projector onto a 
large (macroscopic) subspace. This gives rise to a different set of concerns,
and we refer the reader to the treatment in Ch.~17 of \cite{Grff02c}.


\xb
\section{Applications \label{sct5}}
\xa

Various applications below will illustrate the approach outlined in
Sec.~\ref{sct4}. Those in Secs.~\ref{sbct5.1} and \ref{sbct5.2} show how a
proper application of quantum principles can give physically reasonable results
for the cases considered in Sec.~\ref{sbct2.1} and \ref{sbct2.2}, while
avoiding paradoxes. Simple examples of POVMs and weak measurements are
considered in Secs.~\ref{sbct5.3} and \ref{sbct5.4}. Quantum (non)contextuality
and aspects of the Einstein-Podolsky-Rosen (EPR) paradox are examined in
Secs.~\ref{sbct5.5} and \ref{sbct5.6}.

\xb
\subsection{Spin Half \label{sbct5.1}}
\xa

\xb
\outl{Stern-Gerlach measurement of $S_z$ gives value before the measurement}
\xa

The simplest nontrivial example of a quantum system is the spin of a spin-half
particle, and the spin was first measured in the Stern-Gerlach experiment
mentioned in every textbook.  Using the notation for the eigenstates of the $z$
component of angular momentum $S_z$ introduced earlier in Sec.~\ref{sbct3.1},
suppose that a measurement of $S_z$ corresponds to an isometry
\begin{equation}
 J\ket{z^j} = \ket{\Phi^j}
\label{eqn56}
\end{equation}
of the form \eqref{eqn21}, where $j=+$ or $-$, and the macroscopic outcomes
correspond to projectors $M^+$ and $M^-$ on pointer subspaces satisfying
\eqref{eqn17}. Then \eqref{eqn24} takes the form
\begin{equation}
 [z^+] = J\ad M^+ J,\quad [z^-] = J\ad M^- J.
\label{eqn57}
\end{equation}
Hence if the macroscopic outcome is $M^+$---e.g, an atom is detected in the
upper beam emerging from a Stern-Gerlach magnet---one can conclude using
the family of four histories at times $t_0<t_1<t_2$ (at $t_1$ and $t_2$ choose
one of the two properties inside the curly brackets)
\begin{equation}
  [\psi_0]\ot [\Om_0] \od \{[z^+],[z^-]\} \od \{M^+,M^-\},
\label{eqn58}
\end{equation}
that at time $t_1$ before the measurement began the particle had the property
$[z^+]$ corresponding to $S_z=+1/2$, whatever the initial state
$[\psi_0]$. Similarly, $M^-$ would indicate $S_z=-1/2$ at the earlier time.

\xb
\outl{Probabilities of $[z^\pm]$ before measurement given measurement
  outcomes $M^\pm$}
\xa

One can check this by a direct calculation assuming an initial state
\begin{equation}
 \ket{\psi_0} = \al\ket{z^+} + \bt \ket{z^-},
\label{eqn59}
\end{equation}
and using the chain kets to evaluate the probabilities for the four histories
in \eqref{eqn58}:
\begin{equation}
 \Pr(z^+_1,M^-_2) = \Pr(z^-_1,M^+_2) = 0, \quad
 \Pr(z^+_1,M^+_2) = |\al|^2,\quad  \Pr(z^-_1,M^-_2) = |\bt|^2.
\label{eqn60}
\end{equation}
The marginals and conditionals are then
\begin{equation}
 \Pr(M^+_2) = |\al|^2, \quad \Pr(M^-_2) = |\bt|^2,\quad
\Pr(z^+_1\vbl M^+_2) = 1,\quad \Pr(z^-_1\vbl M^-_2) = 1,
\label{eqn61}
\end{equation}
where the last two hold if $|\al|^2$ (respectively, $|\bt|^2$) is nonzero.
In short, the particle at $t_1$ had the value of $S_z$ indicated by the
measurement outcome at $t_2$, independent of the state $\ket{\psi_0}$ at 
$t_0$, in agreement with \eqref{eqn57}.

\xb
\outl{Earlier $S_x$ basis; $S_z$ measurement outcome tells nothing about $S_x$}
\xa

Next, assuming the same unitary dynamics \eqref{eqn56}, consider a different
family of histories,
\begin{equation}
  [x^+]\ot [\Om_0] \od \{[x^+],[x^-]\} \od \{M^+,M^-\},
\label{eqn62}
\end{equation}
in which the initial $[\psi_0]$ is now $[x^+]$, and the properties at 
$t_1$ refer to $S_x$ instead of $S_z$. It is straightforward to show that
the family is consistent, with joint probabilities (obtained from chain
kets)
\begin{equation}
\Pr(x_1^+,M_2^+) = \Pr(x_1^+,M_2^-) = 1/2,\quad
 \Pr(x_1^-,M_2^+) = \Pr(x_1^-,M_2^-) = 0.
\label{eqn63}
\end{equation}
The conditionals
\begin{equation}
\Pr(x_1^+\vbl M_2^+) = \Pr(x_1^+\vbl M_2^-) = 1,\quad
 \Pr(x_1^-\vbl M_2^+) = \Pr(x_1^-\vbl M_2^-) = 0.
\label{eqn64}
\end{equation}
are exactly the same for $M^+$ and $M^-$, so the measurement outcomes  at $t_2$
tell us nothing at all about $S_x$ at time $t_1$. Instead its value is
determined entirely by the initial state $\ket{x^+}$ at $t_0$.

\xb
\outl{Is $x^+$ or $z^-$ correct answer at $t_1$ given $M^-$ at $t_2$? 
Depends on chosen framework}
\xa

Given the family \eqref{eqn62} and a pointer outcome, say $M^-$ at $t_2$, are
we to infer $S_x=+1/2$ at the earlier time $t_1$ using \eqref{eqn64}, or
$S_z=-1/2$ using \eqref{eqn61}? Both inferences are correct, \emph{but in
  separate frameworks which cannot be combined}. Frameworks are chosen by the
physicist depending on which aspect of the situation is of interest. The
physicist who sets up an apparatus to prepare a spin-half particle with a
particular polarization may wish to explain in quantum mechanical terms how it
functions, in which case the family \eqref{eqn62} is an appropriate starting
point, and \eqref{eqn64} will confirm that later measurements do not have any
undesirable retrocausal influence. On the other hand the physicist who has
constructed an apparatus to measure a particular polarization can best
explain how it functions in that capacity by using the family \eqref{eqn58}.
Even if $[\psi_0]=[x^+]$ is not an eigenstate of $S_z$, \eqref{eqn61} shows that
the later pointer position reveals the prior property the instrument was
designed to measure. These two physicists might be one and the same; several
incompatible frameworks may be useful for analyzing a particular experimental
arrangement, while the single framework rule prevents drawing meaningless
conclusions or paradoxical results.

\xb
\outl{Properties at intermediate time $t_{1.1}$}
\xa

Properties at an additional intermediate time before the measurement has begun,
say $t_{1.1}$, can be added to \eqref{eqn62} to form a consistent family
at times $t_0< t_1<t_{1.1}<t_2$,
\eqref{eqn62}:
\begin{equation}
  [x^+]\ot [\Om_0] \od \{[x^+],[x^-]\}\od \{[z^+],[z^-]\}  \od \{M^+,M^-\},
\label{eqn65}
\end{equation}
where we assume that $T(t_{1.1},t_1) = I$. 
Using it one can show that
\begin{equation}
 \Pr(x^+_1) = 1,\quad
 \Pr(z^+_{1.1} \vbl M^+_2) = \Pr(z^-_{1.1} \vbl M^-_2)=1.
\label{eqn66}
\end{equation}
Thus if the later measurement outcome is $M^-$ one can be sure (based on the
initial state) that $S_x = +1/2$ at $t_1$ and also (based on the measurement
outcome) that $S_z = -1/2$ at $t_{1.1}$. This seems odd if one tries to imagine
a physical process rotating the direction of the spin from $+x$ to $-z$, since
the particle is moving in a field-free region and not subject to a torque. Once
again the \emph{choice of framework} which allows a description of a particular
aspect of the situation must be carefully distinguished from a \emph{dynamical
  physical process}. While there is no exact classical counterpart of a
framework choice, the following analogy may help.. If one looks at a coffee cup
from above one can discern certain things---is it filled with coffee?---which
are not visible from below, whereas things visible from below, such as a crack
in the bottom, may not be visible from above. Changing the point of view does
not change the coffee cup or its contents, but does allow one to see different
things. The analogy with the quantum case breaks down in that it makes sense to
speak of a cup that both contains coffee and has a (small) crack in the bottom,
whereas $S_x=+1/2$ \AND\ $S_z=-1/2$ is meaningless, as the projectors do not
commute. To be sure, $S_x=+1/2$ at an earlier time is correctly combined in
\eqref{eqn65} with $S_z$ at a later time: think of first looking at the coffee
cup from the top and later from the bottom. However, interchanging the
intermediate events in \eqref{eqn65} so that $S_z$ properties at $t_1$ precede
the $S_x$ properties at $t_{1.1}$ results in an inconsistent family. Classical
analogies help, but in the end there is no substitute for a consistent quantum
analysis.

\xb
\subsection{Mach-Zehnder \label{sbct5.2}}
\xa

\xb
\outl{Correspondence between MZ and spin half}
\xa

A correspondence between spin-half measurements as discussed in
Sec.~\ref{sbct5.1} and the Mach-Zehnder setup of Sec.~\ref{sbct2.2} will assist
in understanding the latter. Consider a time $t_1$ at which, see
Fig.~\ref{fgr3}, the photon has been reflected from the upper and lower
mirrors, but has yet to reach the location of the second beam splitter, or, if
the latter is absent, the crossing point of the two trajectories. Let
$\ket{z^+}$ be the part of the photon wavepacket in the upper arm, and
$\ket{z^-}$ the part in the lower arm of the interferometer at this time, and
let $\ket{x^+}$ and $\ket{x^-}$ be the coherent superpositions of $\ket{z^+}$
and $\ket{z^-}$ defined in \eqref{eqn4}. Further assume that the action of the
first beamsplitter in Fig.~\ref{fgr3} is to prepare the photon in the state
$\ket{x^+}$. Let $M^+$ be the projector on the macroscopic subspace in which
$D^+$ in Fig.~\ref{fgr3} has detected the photon while $D^-$ has not, and $M^-$
its counterpart for detection by $D^-$ rather than $D^+$.

\xb
\outl{Second beam splitter absent}
\xa

If the second beamsplitter is absent, Fig.~\ref{fgr3}(b), a photon in the state
$\ket{z^+}$ in the upper arm will trigger $D^+$, while $\ket{z^-}$ in the lower
arm will trigger $D^-$. This can be discussed using a family of four histories
as in \eqref{eqn58}, with $\ket{\psi_0} = \ket{x^+}$:
\begin{equation}
 [x^+]\ot [\Om_0] \od \{[z^+],[z^-]\} \od \{M^+,M^-\}.
\label{eqn67}
\end{equation}
The conditional probabilities are the same as in \eqref{eqn61}: if $D^+$ is
triggered one can be certain the photon was earlier in the state $[z^+]$, so in
the upper arm of the interferometer, whereas detection by $D^-$ indicates the
earlier state $[z^-]$ in the lower arm. These are the same conclusions one
would arrive at from a naive inspection of Fig.~\ref{fgr3}(b), but they have
now been confirmed using an analysis based upon consistent quantum principles.

\xb
\outl{Insert $t_{1.1}$ between $t_1$ and $t_2$. Photon in two arms collapses
  into one?}
\xa

Now add an additional time $t_{1.1}>t_1$ at which the photon is still inside
the interferometer. The consistent family
\begin{equation}
 [x^+]\ot [\Om_0] \od \{[x^+],[x^-]\}\od \{[z^+],[z^-]\}  \od \{M^+,M^-\}
\label{eqn68}
\end{equation}
(where note that histories with $[x^-]$ at $t_1$ have zero probability, so can
be ignored) is formally identical to \eqref{eqn65}, but introduces a new
conceptual difficulty. In the spin-half case the issue was how a spin angular
momentum of $S_x=+1/2$ at $t_1$ could suddenly precess into $S_z=+1/2$ or
$-1/2$ at $t_{1.1}$. However mysterious that might be, one could still imagine
the change taking place at the location of the spin half particle. But for the
Mach-Zehnder $[x^+]$ is a nonlocal superposition between the two arms at $t_1$;
can it suddenly collapse into one or the other arm, $[z^+]$ or $[z^-]$, at a
time $t_{1.1}$, even if the interval between $t_1$ and $t_{1.1}$ is very short,
so making this collapse essentially instantaneous? Is this (seeming)
nonlocality consistent with relativity theory?

\xb
\outl{The ``collapse'' is framework choice, not physical process}
\xa

Just as in the case of spin half this (apparent) paradox may be dealt with by
noting that a change in what is being described is not the same as a physical
process. Thus if the pair $\{[z^+],[z^-]\}$ at $t_{1.1}$ in \eqref{eqn68} is
replaced with $\{[x^+],[x^-]\}$, this new family is again consistent, but the
``collapse'' between $t_1$ and $t_{1.1}$ is no longer present. Families or
frameworks are chosen by the physicist and are not consequences of some law of
nature. See the discussion following \eqref{eqn66}.

\xb
\outl{Superposition of particle in 2 places $\neq$ particle simultaneously 
present in  both}
\xa

In addition it is worth noting that a quantum superposition, such as
$\ket{x^+}$, of a particle at two locations is not at all the same thing as its
being in both places at the same time. Translated into quantum terminology the
statement that the photon is in the upper arm \AND\ in the lower arm becomes
$[z^+]\ \AND\ [z^-]$, which because the projectors commute makes perfectly good
sense, and because their product is zero this conjunction is always false: a
photon can \emph{never} be located simultaneously in both the upper and in the
lower arm, unlike a classical wave.

\xb
\outl{2d beamsplitter + detectors = apparatus to detect $[x^+]$, $[x^+-]$}
\xa

Next consider the case with the second beamsplitter present, and suppose that
the phases are such that a photon in the state $\ket{x^+}$ inside the
interferometer will later be detected by $D^+$, and $\ket{x^-}$ detected by
$D^-$. In this case one can think of the detectors and the the second
beamsplitter as together constituting an apparatus designed to detect $[x^+]$
and $[x^-]$, the photon analogs of the spin-half $S_x$ eigenstates (which for a
spin-half particle could be measured by rotating the Stern-Gerlach apparatus so
that its field gradient is in the $x$ rather than the $z$ direction). The
second beamsplitter changes the unitary dynamics in such a way that
$\ket{z^j}$ on the left side of \eqref{eqn56} is replaced by $\ket{x^+}$, 
and thus \eqref{eqn57} becomes
\begin{equation}
 [x^+] = J\ad M^+J,\quad [x^-] = J\ad M^- J.
\label{eqn69}
\end{equation}
Thus the measurement outcomes now indicate different superposition states of
the photon inside the interferometer; the measurement measures ``which phase?''
rather than ``which path?'', see Fig.~\ref{fgr4}. With the new dynamics
\eqref{eqn67} is no longer a consistent family, but one can instead use
\begin{equation}
 [\psi_0]\ot [\Om_0] \od \{[x^+],[x^-]\} \od \{M^+,M^-\},
\label{eqn70}
\end{equation}
in order to infer from the measurement outcome the presence of one of two
distinct (i.e., orthogonal) superposition states inside the interferometer,
which is to say the difference between photons originating in $S_1$ or 
$S_2$ in Fig.~\ref{fgr4}(a).

\xb
\outl{Paradox of removing (inserting) 2d beamsplitter at the last moment}
\xa

We are now ready to discuss the (supposed) paradox, Sec.~\ref{sbct2.2},
associated with removing or inserting the second beam splitter at the very last
moment just before the photon reaches it. One can think of this change as the
Mach-Zehnder analog of rotating a Stern-Gerlach apparatus about the axis of the
atomic beam just before the arrival of a spin-half particle, so that it will
measure $S_x$ rather than $S_z$. In the absence of this rotation one can use
the measurement outcome to assign a value to $S_z$ before the measurement took
place, whereas if the rotation has taken place before the particle arrives, the
measurement outcome indicates the prior value of $S_x$. This does not mean that
the particle has both an $S_z$ and an $S_x$ value, for these two quantities are
incompatible, and that is why they cannot be measured simultaneously. In the
case of the Mach-Zehnder, if the second beamsplitter is absent the measurement
outcome will indicate either an earlier $[z^+]$ property, photon in the upper
arm, or $[z^-]$, photon in the lower arm. If the second beamsplitter is present
the measurement outcome distinguishes the earlier superposition properties
$[x^+]$ and $[x^-]$, neither of which is compatible with assigning the photon
to one of the arms rather than the other. In neither case is there any need to
suppose that the later measurement choice influences the particle before
measurement. Instead, changing the type of measurement alters what type of
information about the earlier state of the particle can be inferred from the
measurement outcome.


\xb
\subsection{Spin-Half POVM \label{sbct5.3}}
\xa

\xb
\outl{Spin half POVM using 3 nonorthogonal states $\{\ket{u^k}\},\,k=1,2,3$}
\xa

A simple but instructive example of a POVM for a spin-half particle can be
constructed using the three nonorthogonal states
\begin{align}
 \ket{u^1} &= (\ket{z^+} + \ket{z^-})/\st,\quad
 \ket{u^2} = (\om\ket{z^+} + \om^2\ket{z^-})/\st,
\notag\\
 \ket{u^3} &= (\om^2\ket{z^+} + \om\ket{z^-})/\st,\quad
 \om := \exp[2\pi i/3]. 
\label{eqn71}
\end{align}
The projectors $[u^k]$ are associated with points on the equator of the Bloch
sphere: $[u^1]$ at the positive $x$ axis, while $[u^2]$ and $[u^3]$ are
separated from $[u^1]$ and each other by $120\dg$. 
The operators
\begin{equation}
 Q^k := (2/3) [u^k]
\label{eqn72}
\end{equation}
for $k=1,2,3$ sum to the identity and hence constitute a POVM.

\xb
\outl{POVM derived from $J:\HC_s\ra\HC_M$, $d_M=3$. $Q^k = J\ad [k] J =
  (2/3)[u^k]$ }
\xa

This POVM can be obtained from an isometry as discussed in
Sec.~\ref{sbct4.2}, where we assume for simplicity a ``toy'' apparatus Hilbert
space $\HC_M$ of dimension 3, with an
orthonormal basis  $\{\ket{k}\}$, $k=1,2,3$.  The isometry $J$ can be written
in the form
\begin{equation}
 J\ket{u^k} = \ket{v^k} := \sqrt{3/2}\,\ket{k}- \sqrt{1/2}\,\ket{w};\quad
\ket{w} := (\ket{1}+\ket{2}+\ket{3})/\sqrt{3}.
\label{eqn73}
\end{equation}
This $J$ maps the two-dimensional $\HC_s$ into the two-dimensional subspace of
$\HC_M$ consisting of kets orthogonal to $\ket{w}$. The orthogonal measurement
projectors in the notation of  Sec.~\ref{sct4} are:
\begin{equation} 
M^k := [k].
\label{eqn74}
\end{equation}
With the help of the formulas
\begin{equation}
 \ket{u^k} = J\ad \ket{v^k},\quad J\ad\ket{w} =0,\quad
  \ket{k}= \sqrt{2/3}\,\ket{v^k} +\sqrt{1/3}\,\ket{w},
\label{eqn75}
\end{equation}
where the first and second are consequences of $J\ad J=I_s$ and the fact
that $\mted{w}{J}{u^k}=0$, while the third comes from rewriting \eqref{eqn73},
one can show that
\begin{equation}
 J\ad M^k J = J\ad [k] J = (2/3)[u^k] = Q^k,
\label{eqn76}
\end{equation}
in agreement with \eqref{eqn35}.

\xb
\outl{Consistent family with $\{[u^k], I-[u^k]\}\od M^k$ at $t_1,t_2$.}
\xa

The analysis in Sec.~\ref{sbct4.2} shows that the family consisting of
the histories
\begin{equation}
 Y^k = [\psi_0]\ot[\Om_0]\od \{[u^k],I_s-[u^k]\}\od M^k
\label{eqn77}
\end{equation}
at times $t_0 < t_1 < t_2$, $k=1,2,3$, is consistent for any initial spin-half
state $\ket{\psi_0}$. Note that the PDI $\{[u^k],I_s-[u^k]\}$ at $t_1$ is
linked to the final $M^k$, and because the $[u^k]$ are not orthogonal these
intermediate PDIs for different $k$ are incompatible. This is not a problem,
because in a particular run only one measurement outcome corresponding to a
specific $k$ will occur, and for \emph{that} $k$ one can be sure (see the
discussion in Sec.~\ref{sbct4.2}; here $Q^k$ is proportional to a rank-one
projector) the particle was at time $t_1$ in the state $[u^k]$, since the
history with the event $I_s-[u^k]$ has zero weight.

\xb
\outl{Intermediate time properties $\{[u^k], I-[u^k]\}$ depend on final $M^k$
  $\not\Rightarrow$ retrocausation}
\xa

That the framework used to describe the situation at $t_1$ depends on the later
measurement outcome at time $t_2$ should not be misunderstood. It is not the
case that a later event \emph{caused} an earlier one. Rather, a specific later
outcome of a process which is intrinsically random allows one to reach a
conclusion which would not have been possible had the outcome been different.
There are classical analogs of this, though of course they all
have limitations when discussing quantum systems. One should also
keep in mind that while the family \eqref{eqn77} provides a rather natural
interpretation of the measurement outcome $k$, the choice is not unique.

\xb
\outl{Calibration of POVM apparatus using $[u^k]$ or $I-[u^k]$}
\xa

It is worth considering what happens if one is trying to calibrate the POVM
apparatus using several runs in which particles are prepared in the state
$[u^1]$, i.e., $S_x = 1/2$. The probability of outcome $k$ will be
\begin{equation}
 \Pr(M^k_2) = \Tr([u^1]\, Q^k) = \begin{cases}
 2/3 &\text{if $k=1$}\\
 1/3 &\text{if $k=2$ or $3$}. \end{cases}
\label{eqn78}
\end{equation}
Thus unlike the situation for a PDI, the prior preparation does not determine
the measurement outcome, although $M^1$ is more likely to occur than either of
its alternatives. If for some run the outcome is $M^2$ we might conclude, using
\eqref{eqn77} with $k=2$, that the particle was earlier in the state $[u^2]$,
even though we know it was prepared in $[u^1]$. This is not a paradox as long
as one remembers that quantum theory allows the use of different frameworks,
and one is careful not to combine incompatible frameworks in violation of the
single framework rule.
An alternative calibration procedure uses particles
prepared in states orthogonal to the $[u^k]$. For example, $[x^-]$ is orthogonal
to $[u^1]$, and if in \eqref{eqn77} $\ket{\psi_0} = \ket{x^-}$, the outcome 
probabilities are:
\begin{equation}
 \Pr(M^k_2) = \Tr(\,[x^-]\, Q^k) = \begin{cases}
 0 &\text{if $k=1$}\\
 1/2 &\text{if $k=2$ or $3$}. \end{cases}
\label{eqn79}
\end{equation}
The fact that in this case the $k=1$ outcome is never observed is an
indication that the apparatus is functioning properly. 


\xb
\subsection{Weak Measurements\label{sbct5.4}}
\xa

\xb
\outl{Definition. Can use probe later subjected to a strong measurement}
\xa

\xb
\outl{Variations: Successive weak measurements. Post selection}
\xa

A \emph{weak measurement} is one in which the measured system, the particle,
interacts very weakly with the measurement apparatus. As a consequence a single
measurement provides very little information about the particle, so weak
measurements are usually employed in a situation in which the measurement can
be repeated many times, each time with a particle prepared in the same state
before the measurement. One way of implementing a weak measurement is to let
the particle interact weakly with a another microscopic system, called a probe,
which has itself been prepared in a known quantum state. After interacting with
the particle the probe is subjected to a projective (``strong'') measurement by
a macroscopic apparatus, with the intent of learning something about the
original particle in this indirect way.
There are many possible variations of this procedure. For example, the same
particle may be subjected to a succession of weak measurements, one after the
other, each supplying some additional information. Or, after interacting with
the probe, the particle may itself be subjected to a strong measurement. When
attention is focused on cases resulting in some particular outcome of the final
strong measurement on the particle one speaks of \emph{post selection}. There
is an enormous literature on weak measurements; for access to some of it see
\cite{TmCh13,Svns13,Wkpd17}.

\xb
\outl{Weak values: controversial, not associated with physical properties}
\xa

Weak measurements have no necessary connection with \emph{weak values}, though
the two are often discussed together, and sometimes it is assumed that weak
measurements can or should be be interpreted using weak values. The physical
significance of weak values has been the subject of an ongoing controversy
\cite{Svns13}. Suffice it to say that in general there is no reason to think of
a weak value as linked to a physical property, or the average value of a
physical variable, at least as those terms are employed in the present article,
where they are associated with Hilbert subspaces.

\xb
\outl{Weak measurement, or weak + strong = POVM; applied to example}
\xa

\xb
\outl{Particle in $\ket{A},\ket{B}$; probe in
$\ket{j},\,j=0,1,2$: probability $\ep$ for $\ket{A0}\ra\ket{A1}$,
$\ket{B0}\ra\ket{B2}$}
\xa

A weak measurement, either by itself or when followed by a strong measurement,
can be understood as a particular type of POVM, and thus understood in terms of
prior properties of the particle as discussed in Sec.~\ref{sbct4.2}. The
following simple example illustrates how this works in a particular case. Let
the particle be a two-state system with an orthonormal basis $\{\ket{A},
\ket{B}\}$, which one can think of as representing a photon in one of two
channels, as in the Mach-Zehnder interferometer considered earlier in
Sections~\ref{sbct2.2} and \ref{sbct5.2}. The three-dimensional Hilbert space
$\HC_r$ for the probe has an orthonormal basis $\{\ket{j}\}, j=0,1,2$. We
assume the probe is initially in the state $\ket{0}$, while the particle is in a
superposition
\begin{equation}
 \ket{\psi_0} = a\ket{A} + b\ket{B}.
\label{eqn80}
\end{equation}
The particle-probe interaction results in a unitary time development 
\begin{equation}
 T(t_2,t_1) \bigl(\,\ket{A}\ot\ket{0}\bigr) = 
\ket{A}\ot(\zt\ket{0}+\eta\ket{1}),\quad
 T(t_2,t_1) \bigl(\,\ket{B}\ot\ket{0}\bigr) = 
 \ket{B}\ot(\zt\ket{0}+\eta\ket{2}),
\label{eqn81}
\end{equation}
during the interval from $t_1$ to $t_2$ (or $t_0$ to $t_2$, given our usual
assumption that $T(t_2,t_0) = T(t_2,t_1)$), where
\begin{equation}
 \zt = \sqrt{1-\ep},\quad \eta = \sqrt{\ep}.
\label{eqn82}
\end{equation}
Here $\ep$, a measure of the strength of the particle-probe interaction, is
assumed to be very small, so that the probability is high that the probe will
be left in its initial untriggered state $\ket{0}$, but on rare occasions it
will be kicked to $\ket{1}$ if the particle is in channel $A$, or to
$\ket{2}$ if the particle is in $B$. Feynman's use of a weak light source
in Sec.~1-6 of \cite{FyLS651} is a good illustration of this idea.

\xb
\outl{Particle measured during $t_2\ra t_3$ in basis $\ket{E},\ket{F}$; probe
in basis $\ket{j},j=1,2,3$}
\xa

After this, during the time interval from $t_2$ to $t_3$ the probe is measured
in the $j=0,1,2$ basis, and the particle is measured in an orthonormal basis
$\ket{E},\ket{F}$ related to $\ket{A},\ket{B}$ by
\begin{equation}
 \ket{A} = \al_e\ket{E}+\al_f\ket{F},\quad 
 \ket{B} = \bt_e\ket{E}+\bt_f\ket{F},
\label{eqn83}
\end{equation}
where 
\begin{equation}
 \mat{ \al_e & \al_f\\ \bt_e & \bt_f}
\label{eqn84}
\end{equation}
is a unitary matrix. Different choices of these parameters could be used to
represent different situations analogous to those shown in Fig.~\ref{fgr3},
where the second beam splitter is either present or absent.

\xb
\outl{Isometries $J_s,\,J_r$ combined yield $J:\HC_s \ra$ measurement outputs}
\xa

As the particle and the probe are measured by separate devices we can
associate with each an isometry mapping from $t_2$ to $t_3$:
\begin{equation}
 J_s\ket{E} = \ket{\Phi_s^E},\quad
 J_s\ket{F} = \ket{\Phi_s^F}, \quad
 J_r\ket{j} = \ket{\Phi_r^j} \text{ for $j=1,2,3$}.
\label{eqn85}
\end{equation}
Combining these with \eqref{eqn81} yields an isometry mapping $\HC_s$ at time
$t_1$ to both outputs at time $t_3$:
\begin{align}
 J\ket{A} &= \zt\bigl( \al_e\ket{\Phi^{E0}} +\al_f\ket{\Phi^{F0}}\bigr) +
 \eta \bigl( \al_e\ket{\Phi^{E1}} +\al_f\ket{\Phi^{F1}}\bigr),
\notag\\
 J\ket{B} &= \zt\bigl( \bt_e\ket{\Phi^{E0}} +\bt_f\ket{\Phi^{F0}}\bigr) +
 \eta \bigl( \bt_e\ket{\Phi^{E2}} +\bt_f\ket{\Phi^{F2}}\bigr),
\label{eqn86}
\end{align}
where $\ket{\Phi^{E0}}$ is shorthand for $\ket{\Phi_s^E}\ot\ket{\Phi_r^0}$, and
lies in the subspace $M^{E0} = M^E\ot M^0$ for the two pointers, and similarly
for the other cases.
 
\xb
\outl{$J\ad(\cdot)J$ applied to pointer projectors $\ra Q^{E0}$, etc.}
\xa

The backward $J\ad(\cdot)J$ map applied to the pointer projectors yields POVM
elements which are operators on $\HC_s$ and thus can be written as $2 \tm 2$
matrices in the $\ket{A}, \ket{B}$ basis:
\begin{align}
 Q^{E0} = (1-\ep)\mat{|\al_e|^2 & \al_e^* \bt_e\\ \al_e \bt_e^* &|\bt_e|^2},&
\quad
 Q^{F0} = (1-\ep)\mat{|\al_f|^2 & \al_f^* \bt_f\\ \al_f \bt_f^* &|\bt_f|^2},
\notag\\
 Q^{E1} = \ep\mat{|\al_e|^2 & 0 \\ 0 & 0},\quad
 Q^{E2} = \ep\mat{0 & 0 \\ 0 & |\bt_e|^2},&\quad
 Q^{F1} = \ep\mat{|\al_f|^2 & 0 \\ 0 & 0},\quad
 Q^{F2} = \ep\mat{0 & 0 \\ 0 & |\bt_f|^2}.
\label{eqn87}
\end{align}
That these six operators sum to the identity follows from the unitarity of
\eqref{eqn84}: its rows are orthogonal, so $\al_e^* \bt^{}_e + \al_f^* \bt_f^{}
=0$, and its column vectors are normalized.

\xb
\outl{Physical interpretation of POVM elements $Q^{E0}$, etc.}
\xa

The simple form of the last four matrices in \eqref{eqn87} can be understood by
noting that the probe, which starts off in $\ket{0}$, can reach state $\ket{1}$
only if the particle is in channel $A$, which is why $Q^{E1}$ and $Q^{F1}$ are
proportional to the projector $[A]$; similarly, only if the particle is in $B$
can the probe arrive at $\ket{2}$. The more complicated matrix $Q^{E0}$ is
$1-\ep$ times the projector $[E]$, which is reasonable since in this case the
probe was not triggered but remained in $\ket{0}$, so did not perturb the
particle; similarly, $Q^{F0}=(1-\ep)[F]$.

\xb
\outl{We used Sec.~\ref{sbct4.2} to find measured properties; unlike
  textbook approach}
\xa

Our discussion has employed the strategy introduced in Sec.~\ref{sbct4.2}, of
interpreting outcome $k$ of a generalized measurement in terms of properties
that correspond to diagonalizing the operator $Q^k$. In this example each $Q^k$
is proportional to a pure state projector, so the interpretation is relatively
simple, and is independent of the initial state $\ket{\psi_0}$ of the particle
in \eqref{eqn80}. The \emph{probabilities} of various measurement outcomes will
depend upon the coefficients $a$ and $b$ in $\ket{\psi_0}$, and can be computed
from the POVM matrices using $\mte{\psi_0}{Q}$, whereas the \emph{physical
  interpretation} of each outcome in terms of prior properties does not depend
on $\ket{\psi_0}$. The framework used here is convenient for discussing what a
quantum measurement measures, but does not exclude the use of other frameworks.
The standard textbook computational procedure uses the entangled state
$T(t_2,t_1)(\ket{\psi_0}\ot\ket{0})$ to calculate probabilities of various
measurement outcomes, and there is nothing wrong with that when one is only
interested in those probabilities, and not in how these outcomes reveal the
properties of the particle that the apparatus was designed to measure.

\xb
\outl{Reference to RBG nested MZ paper for a more complicated situation}
\xa

For an analogous discussion (without using the language of POVMs) of a
more complicated situation, which has given rise to some controversy, see
Sec.~V of \cite{Grff16}.



\xb
\subsection{Is Quantum Mechanics Contextual? \label{sbct5.5}}
\xa

\xb 
\outl{`Contextual' in claim `QM is contextual' is unclear. Define
  `Bell contextual' by: $[A,B]=[A,C]=0; [B.C]\neq 0$; $A$ value is
  different when $A$ measured with $B$, or with $C$} 
\xa

One often encounters the claim that ``quantum mechanics is contextual''%
\footnote{Some authors make it clear that it is \emph{hidden variables}
  versions of quantum mechanics which are contextual, but many omit that
  qualification; for a recent (but hardly unique) example, see \cite{Smao17}.}.
Unfortunately the term ``contextual'' is used in more than one way. A
relatively precise definition due to Bell \cite{Bll661} and used in some
later quantum foundations literature, e.g., Sec.~VII of \cite{Mrmn93} and
p.~188 of \cite{Prs93} is the following: Let $A$, $B$, and $C$ be three
observables (i.e., Hermitian operators), and suppose that $A$ commutes with $B$
and $C$, but $B$ and $C$ do not commute:
\begin{equation}
 [A,B]=0,\quad [A,C]=0,\quad [B,C] \neq 0.
\label{eqn88}
\end{equation}
This means that $A$ can (in principle) be measured together with $B$, or
together with $C$, whereas $B$ and $C$ are incompatible and cannot be measured
together. One can then ask: does the measured value of $A$ depend on whether it
is measured together with $B$ or with $C$? If the answer is ``yes,'' then
quantum mechanics (or whatever theory is being discussed) is \emph{contextual},
and if ``no,'' it is \emph{noncontextual}. To avoid confusion, let us add a
modifier and refer to \emph{Bell (non)contextual} when these terms are used in
the way just described. The following argument will show that quantum mechanics
in the consistent histories interpretation is \emph{Bell noncontextual}. (A more
recent and somewhat different definition of ``contextual'' is discussed briefly
at the end of this section.) 

\xb
\outl{Device with switch: $\bt\ra$ measure $A,B$; $\gm\ra$ measure $A,C$.
  Calibration}
\xa

The definition given above runs into the following difficulty. In a single
experimental run $A$ cannot be measured together with both $B$ and $C$, since
$B$ and $C$ cannot be measured in the same run. And the measured value of $A$
may vary randomly from run to run, making it difficult to make a comparison
between those in which $A$ is measured with $B$ and those in which it is
measured together with $C$. Let us explore this difficulty by thinking of an
apparatus equipped with a switch with two settings: $\bt$ and $\gm$. With the
switch at $\bt$ the apparatus will measure both $A$ and $B$; while with the
setting $\gm$ it will measure $A$ and $C$. We suppose that the apparatus has
been calibrated, Sec.~\ref{sbct2.3}, for $A$ measurements for both switch
settings, so the experimenter can be reasonably confident that the $A$ pointer
outcome will give the correct answer if the input state is an eigenstate of
$A$. Similarly, $B$ calibrations can be carried out with the switch at
$\bt$, and $C$ calibrations for the $\gm$ setting.

\xb
\outl{A=a outcome given $\bt$; same would have occurred in case of $\gm$} 
\xa

Now we ask: suppose that with the $\bt$ setting the $A$ measurement outcome
corresponds to a particular eigenvalue, say $a_3$. Would this outcome have
been the same if the switch setting had been $\gm$? Counterfactual questions of
this sort are a bit tricky; see \cite{Grff99} and Ch.~19 of \cite{Grff02c} for
a proposal that gives plausible results in a quantum setting. For the present
discussion the basic idea is that if one can reliably infer from the apparatus
outcome with switch setting $\bt$ the eigenvalue of $A$ that characterized the
particle \emph{before} any interaction with the apparatus it seems reasonable
that changing the switch from $\bt$ to $\gm$ at the very last moment could not
have altered that earlier property, so the result would have been the same with
the $\gm$ setting, given that the apparatus had been calibrated.

\xb
\outl{3-state model,  $A$, $B$, $C$ with $[A,B]=0=[A,C]$; $BC\neq CB$}
\xa

To make things less abstract, consider a spin-one particle, and let
$\ket{1},\,\ket{2},\,\ket{3}$ be an orthonormal basis for its Hilbert space
$\HC_s$. Define the following observables using dyads:
\begin{equation}
 A = \dya{1} - \dya{2}-\dya{3},\quad
 B = \hf\dya{1} +\dya{2} -\dya{3},\quad
 C = 2\dya{1} +\dyad{2}{3} + \dyad{3}{2}.
\label{eqn89}
\end{equation}
It is obvious that $[A,B] = 0$, and straightforward to show that $[A,C] = 0$
and $[B,C]\neq 0$.

\xb
\outl{Fig. 5: apparatus for $A$ with $B$ or $A$ with $C$ measurement}
\xa

\begin{figure}[h]
$$\includegraphics{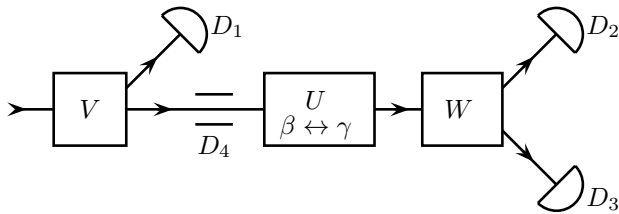}$$
\caption{Apparatus to measure $A$ along with $B$ ($U_\bt$), or with $C$ 
($U_\gm$).}
\label{fgr5}
\end{figure}

\xb
\outl{Description of measurement process}
\xa

A possible apparatus for measuring these observables is shown schematically in
Fig.~\ref{fgr5}. The incoming particle first passes through a device $V$ (one
can think of an electric field gradient acting on a particle with an electric
quadrupole moment) which splits the path in two. The upper path is followed by
a particle in the state $\ket{1}$ and leads to the detector $D_1$. The lower
(straight) path is followed by a particle whose state is any linear
combination of $\ket{2}$ and $\ket{3}$, and it passes through a nondestructive
detector $D_4$ that measures the particle's passage without disturbing its
internal state. Following this there is another device $U$ with a switch: if
the switch setting is $\bt$ it carries out a unitary transformation $U_\bt$
equal to the identity $I$ (i.e, the device does nothing), while if the setting
is $\gm$ the unitary is
\begin{equation}
 U_\gm = (1/\st)\bigl\{\dya{2} + \dyad{2}{3} + \dyad{3}{2}  -\dya{3}\bigr\}.
\label{eqn90}
\end{equation}
Then yet another device $W$ (think of a Stern-Gerlach magnet)
splits the trajectory into one moving upwards if the particle state is
$\ket{2}$, or downwards if it is $\ket{3}$; these terminate in
detectors $D_2$ and $D_3$.

\xb
\outl{What happens if incoming particle is in eigenstate of A or of B or of C}
\xa

\xb 
\outl{Switch setting $\bt\ra\gm$ made later cannot alter $A$ measurement
  outcome} 
\xa

A particle initially in the eigenstate $\ket{1}$ of $A$ with eigenvalue $+1$
will be detected by $D_1$, whereas any eigenstate of $A$ with eigenvalue $-1$,
i.e., any linear combination of $\ket{2}$ and $\ket{3}$, will be detected by
$D_4$ and then travel on. Thus a measurement of $A$ precedes the particle's
passing through the box $U$, and the outcome will not be affected by whether
the unitary is $U_\bt$ or $U_\gm$. The switch setting could, in principle, be
decided at the very last moment, after the particle (if on this path) has
passed through $D_4$. A measurement of $B$ is carried out by setting $U=I$, so
that initial eigenstates with eigenvalues of 1/2, 1, and $-1$ will be detected
by detectors 1, 2, and 3, respectively. Alternatively, $C$ can be measured by
setting $U=U_\gm$, \eqref{eqn90}, and its eigenvalues of 2, 1, and $-1$
correspond to detection by detectors 1, 2, and 3, respectively. It should be
clear from the construction shown in Fig.~\ref{fgr5} that if the change from an
$A$-plus-$B$ apparatus to an $A$-plus-$C$ apparatus, by moving the switch from
$\bt$ to $\gm$, is made \emph{after} the particle has passed the position of
detectors $D_1$ and $D_4$, this cannot affect the $A$ measurement outcome,
assuming the future does not influence the past. Thus in this case it seems
evident that the measurement is (Bell) noncontextual.

\xb
\outl{Generalization from 3 states to arbitrary number of states}
\xa

The preceding discussion for a particle with three states is easily generalized
to the case of an arbitrary (finite) number of states. To see this, let
$\{P^j\}$ be the PDI that diagonalizes $A$ with different projectors
associated with different eigenvalues; i.e.,
\begin{equation}
 A = \sum_\al a_j P^j,
\label{eqn91}
\end{equation}
and $a_j\neq a_{j'}$ when $j\neq j'$. Then it is straightforward to show that
if $A$ commutes with $B$, every $P^j$ in \eqref{eqn91} also commutes with $B$.
So if a basis is chosen such that the matrix of $A$ is diagonal with separate
blocks for each eigenvalue, the matrix of $B$ will be block diagonal, and each
of its blocks can be separately diagonalized by a change of basis that leaves
the $A$ matrix unchanged. The same comment applies to any other observable $C$
that commutes with $A$, whether or not it commutes with $B$, though of course
the bases used to diagonalize $B$ and to diagonalize $C$ must be different if
$[B,C]\neq 0$. The $V$ box in Fig.~\ref{fgr5} separates incoming particles into
separate beams corresponding to the different eigenvalues of $A$, and in
each beam there is a nondestructive detector that plays the role of $D_4$ in
Fig.~\ref{fgr5}. These measurements determine the value of $A$. Next in each
beam there is a unitary that depends on the choice of $\bt$ or $\gm$, followed
by a final set of detectors from which the eigenvalues of $B$ or $C$, as the
case may be, can be inferred.

\xb
\outl{What if $B$ or $C$ measured \emph{before} $A$? General $\ket{\psi_0}$ 
input?}
\xa

\xb
\outl{Use $P^k = J\ad M^k J$ to arrive at 
$\Pr(P_1^j \vbl M_2^k) = \dl_{jk}$}
\xa

This example leaves open the possibility that if the time ordering were
different, $B$ or $C$, as the case might be, measured \emph{before}
$A$, this might have an effect on the value of $A$. Also we have been assuming
that the particle enters the apparatus in an eigenstate of $A$; what if it is
in some arbitrary superposition state $\ket{\psi_0}$? Both concerns are easily
handled using the measurement model introduced in Sec.~\ref{sct4}. In
particular, \eqref{eqn35} takes the form
\begin{equation}
 P^k = J\ad M^k J,
\label{eqn92}
\end{equation}
for a projective measurement associated with the PDI $\{P^k\}$, the obvious
generalization of \eqref{eqn24}. Thus one can be certain that the particle
possessed the property $P^k$ corresponding to the eigenvalue $a_k$ of $A$ at
the time $t_1$ before the measurement took place, given the later measurement
outcome (pointer position) $k$ that corresponds to $M^k$. What went on at an
intermediate time cannot alter this, always assuming the apparatus has been
properly calibrated, so that \eqref{eqn92} holds. Hence quantum measurements
carried out with a properly designed and tested apparatus are noncontextual,
and in this sense quantum theory is (Bell) \emph{noncontextual}.

\xb
\outl{Reasons why QM often considered contextual. 1. Use of Cl HVs by Bell}
\xa

So why is it that one is sometimes told, often with great confidence, that
quantum theory is \emph{contextual}? Various reasons suggest themselves. The
first is that measurements are not properly treated in textbooks. One admires
textbook authors (e.g., \cite{Lloe01,Lloe12,Wnbr15}) who are brave enough to
agree publicly with Bell \cite{Bll901}: they have not been able to solve the
measurement problem. And without some, at least implicit, theory of quantum
measurements one cannot even begin to discuss contextuality in Bell's sense of
the word. Another reason is that in attempting to fill this serious gap in the
textbooks, John Bell and others have proposed that microscopic properties
rather than being represented by Hilbert subspaces might correspond to hidden
variables which in certain crucial respects are \emph{classical}. This is
obvious in the best-known hidden variables approach, the de Broglie-Bohm pilot
wave \cite{BcVl09,Glds12}, where a quantum particle is assumed to have a
well-defined classical position at all times. But it is also true of the
mysterious quantity $\lm$ that appears in many discussions of Bell
inequalities. There is always an assumption of classical behavior on the part
of this mythical object, as has been pointed out repeatedly by Fine, e.g.,
Sec.~3 of \cite{Fne82}, and clearly comes to light in a proper quantum
mechanical analysis of the situation \cite{Grff11}. Even when authors declare
that $\lm$ is or could be the ``quantum state,'' they are not referring to the
noncommuting projectors representing quantum properties in von Neumann's sense.
Decades of research on hidden variables theories have not come close to solving
the second measurement problem \cite{Spkk07,Lfr14,JnLf15}.

\xb
\outl{2. KS and Mermin square paradoxes as arguments $\Ra$ QM is contextual}
\xa

Sometimes the paradoxes and associated inequalities of Kochen and Specker
\cite{KcSp67}, the Mermin square (Sec.~V of \cite{Mrmn93}), and the like are
invoked as grounds for believing that quantum mechanics is contextual, so it is
worth pointing out where such claims go astray, at least in the case of what we
are calling Bell contextuality. (For a more detailed discussion, see Ch.~22 of
\cite{Grff02c}.) Suppose $A$ commutes with $B$. Then, see the discussion
following \eqref{eqn91}, it is possible to write down a collection of pairs of
eigenvalues $(a_j,b_k)$, each pair corresponding to some well-defined and
nontrivial (i.e., not just the zero vector) Hilbert subspace where $A$ takes
the value $a_j$ and $B$ the value $b_k$. Similarly, if $A$ commutes with $C$
one can construct a similar list $(a_j,c_l)$ of possible joint values. One
might suppose that by comparing these two lists one could find pairs
$(b_k,c_l)$ of possible joint values of $B$ and $C$. In particular, suppose
that $(a_2,b_2)$ is a member of the first list. Then there would surely be at
least one pair in the second list, say $(a_2,c_3)$, such that $(b_2,c_3)$ is a
pair of possible simultaneous values for $B$ and $C$. Perfectly good classical
reasoning, but it can fail in the quantum case if $B$ and $C$ do not commute;
the reader can construct an example using \eqref{eqn89}. By applying this
reasoning, which violates the single framework rule, a sufficient number of
times using a sufficient number of observables one can arrive at a
contradiction, and this, so it is claimed, implies that quantum mechanics is
contextual. But this is not a demonstration of the Bell contextuality of
quantum mechanics; instead it shows that the single framework rule must be
taken seriously if one wishes to reason in a consistent way about microscopic
quantum systems.

\xb
\outl{Were QM contextual that would undermine it as an experimental science}
\xa

It is worth remarking that if Bell contextuality were true this would seriously
undermine quantum physics as an experimental science, since experimenters often
interpret their data in terms of prior microscopic properties once the
apparatus has been calibrated. And calibration refers to the quantity of
interest, $A$ in the above discussion, not to other observables which the
apparatus might quite incidentally be measuring at the same time. It would be
an insuperable task to take all of these other possibilities into account when
designing or calibrating equipment. Thus experimental physics relies upon the
fact that quantum mechanics is Bell \emph{noncontextual}.

\xb
\outl{Alternative `contextual' definition found in the literature}
\xa

Finally, there is an alternative definition of ``contextual'' that appears to
underlie many of the more recent discussions in the literature, and receives a
precise definition in \cite{AbBM17}. A \emph{context} is defined to be a
collection of commuting observables which can be measured simultaneously, thus
associated with a single PDI, or in consistent histories terminology a
framework. In the example in \eqref{eqn89}, $A$ and $B$ belong to one context,
and $A$ and $C$ to another, but there is no context (framework) which contains
all three. Given some collection of contexts and a single initial quantum
state, one can use the Born rule to compute the probabilities of measurement
outcomes for operators in each context. The probability assigned to a
particular operator $A$ that belongs to several different contexts is
independent of the context (as expected, since quantum theory is Bell
noncontextual). However there may not exist a \emph{joint} probability
distribution for the \emph{entire} collection of observables if not all of them
commute, and hence there is no single context that contains them all. The
\emph{absence} of such a joint distribution is taken to indicate that quantum
mechanics (or whatever theory is under consideration) is \emph{contextual}.
Perhaps ``multicontextual'' would be a better term. 


\xb
\subsection{Einstein-Podolsky-Rosen-Bohm\label{sbct5.6}}
\xa

\xb
\outl{Bohm version of EPR}
\xa

The Einstein-Podolsky-Rosen (EPR) paradox \cite{EnPr35} is well known and has
given rise to an enormous number of publications. The purpose of the following
remarks is to relate it to the second measurement problem, using Bohm's
simple version of EPR in \cite{Bhm51s}. It makes use of the singlet spin state
\begin{equation}
 \st\,\ket{\psi_0} = \ket{z^+}_a\ot  \ket{z^-}_b - \ket{z^-}_a\ot  \ket{z^+}_b
 =\ket{x^+}_a\ot  \ket{x^-}_b - \ket{x^-}_a\ot  \ket{x^+}_b
\label{eqn93}
\end{equation}
in the Hilbert space $\HC_a\ot\HC_b$ of two spin-half particles $a$ and $b$,
thought of as quite far apart so they do not interact with each other, and
particle $b$ will not interact with an apparatus carrying out a measurement on
particle $a$.

\xb
\outl{EPR paradox stated for Bohm version}
\xa

The essence of the original EPR argument expressed using Bohm's model is as
follows. A measurement of $S_z$ for particle $a$ can be used to infer the value
of $S_z$ for $b$, and since particle $a$ and the apparatus are not interacting
with $b$, that particle must have had that value of $S_z$ before the
measurement of $a$ took place. The property of particle $b$ was, so-to-speak,
``really there,'' a part of physical reality. But one could just as well
measure $S_x$ for particle $a$, and via the same sort of argument assign a
value to $S_x$ for particle $b$, which again would be ``really there.'' But in
the two-dimensional Hilbert space of a spin-half particle there is nothing to
represent a situation in which both $S_x$ and $S_z$ simultaneously take on
particular values. Thus the Hilbert-space approach does not provide a
complete description of physical reality; something is missing.

\xb
\outl{Measurement of $S_z$ of particle $a$ (no measurement of particle $b$)}
\xa

We shall assume that only particle $a$  is measured, and that since neither it
nor the measurement apparatus can interact with particle $b$, the corresponding
isometry $J$, see Sec.~\ref{sbct4.1}, that relates the spin states of both
particles, $\HC_s = \HC_a\ot\HC_b$, to the measurement outcome can be written
in the form:
\begin{equation}
  J \bigl(\ket{\psi}_a\ot \ket{\chi}_b\bigr) = 
 \bigl(J_a \ket{\psi}_a\bigr) \ot \ket{\chi}_b,
\label{eqn94}
\end{equation}
where $\ket{\psi}$ and $\ket{\chi}$ are any two elements of $\HC_a$ and
$\HC_b$, and $J_a:\HC_a \ra \HC_M$ is the isometry for a measurement of
particle $a$ alone. For an $S_z$ measurement, $J_a$ tales the form
\begin{equation}
 J_a\ket{z^+}_a = \ket{A^+},\quad J_a\ket{z^-}_a = \ket{A^-},\quad
M^+ \ket{A^+}=\ket{A^+},\quad M^- \ket{A^-} = \ket{A^-},
\label{eqn95}
\end{equation}
where, as in Sec.~\ref{sbct4.1}, $M^+$ and $M^-$ are projectors on the
macroscopic pointer position subspaces representing the possible outcomes of the
measurement. The counterpart of \eqref{eqn24} is
\begin{equation}
  [z^k]_a \ot I_b = J\ad \bigl(M^k\ot I_b\bigr) J,\quad k = + \text{ or } -,
\label{eqn96}
\end{equation}
where $I_b$ is the identity for particle $b$.

\xb
\outl{Histories with $S_z$ for $a$, $b$ at intermediate time $t_1$}
\xa

Consider a family of histories at times $t_0<t_1<t_2$:
\begin{equation}
 [\psi_0]\ot [\Om_0] \od \{\,[z^+]_a\,,[z^-]_a\}\ot \{\,[z^+]_b\,,[z^-]_b\} \od 
 \{M^+,M^-\},
\label{eqn97}
\end{equation}
where the four projectors $[z^+]_a\ot [z^+]_b$, etc., at the intermediate time
sum to the identity on $\HC_a\ot\HC_b$. There are eight histories in this
family, but we only need to pay attention to those in which $[z^+]_a$ at time
$t_1$ is followed by $M^+$ at $t_2$, or $[z^-]_a$ by $M^-$, since the other
chain kets vanish. But in addition, for $\ket{\psi_0}$ as defined in
\eqref{eqn93},
\begin{equation}
 \bigl(\,[z^+]_a\ot[z^+]_b\,\bigr)\,\ket{\psi_0} =
 \bigl(\,[z^-]_a\ot[z^-]_b\,\bigr)\,\ket{\psi_0} = 0.
\label{eqn98}
\end{equation}
This means that only two histories have positive probabilities:
$[z^+]_a\ot[z^-]_b$ followed by $M^+$ or $[z^-]_a\ot[z^+]_b$ followed by $M^-$.
The chain kets are obviously orthogonal, so the family is consistent, and each
of these histories is assigned a probability of 1/2, leading to the conditional
probabilities:
\begin{equation}
 \Pr(\,z^+_{a1}\vbl M^+_2) = \Pr(\,z^-_{b1}\vbl M^+_2)  = 1,\quad
 \Pr(\,z^-_{a1}\vbl M^-_2) = \Pr(\,z^+_{b1}\vbl M^-_2) = 1.
\label{eqn99}
\end{equation}
In words, the outcome $M^+$ of the measurement of $S_z$ for particle $a$
indicates that at the earlier time $S_z$ was $+1/2$ for particle $a$ and $-1/2$
for particle $b$, while $M^-$ means $S_z$ was $-1/2$ for $a$ and $+1/2$ for $b$.

\xb
\outl{Histories with $S_x$ for $a$, $b$ at $t_1$ when $S_z$ for $a$ is measured}
\xa

A second consistent family, using the same isometry \eqref{eqn95} appropriate
for measuring $S_z$, employs
eigenstates of $S_x$ rather than $S_z$ at $t_1$:
\begin{equation}
 [\psi_0]\ot [\Om_0] \od \{\,[x^+]_a\,,[x^-]_a\}\ot \{\,[x^+]_b\,,[x^-]_b\} \od 
 \{M^+,M^-\}.
\label{eqn100}
\end{equation}
In this case the chain kets in which $[x^+]_a$ and $[x^-]_a$ are followed by
$M^+$ or $M^-$ do not have to vanish. However, the initial $[\psi_0]$
eliminates histories that contain $[x^+]_a\ot [x^+]_b$ or $[x^-]_a\ot [x^-]_b$
at $t_1$, leaving only four nonzero chain kets, which are orthogonal (something
the reader may wish to check). The resulting probabilities then lead to:
\begin{align}
 \Pr(x^+_{a1}\ot x^-_{b1}\vbl M^+_2)
&=\Pr(x^+_{a1}\ot x^-_{b1}\vbl M^-_2) = 1/2 
\notag\\
  \Pr(x^-_{a1}\ot x^+_{b1}\vbl M^+_2)
&=\Pr(x^-_{a1}\ot x^+_{b1}\vbl M^-_2)  = 1/2. 
\label{eqn101}
\end{align}
Since these conditional probabilities are the same for the two measurement
outcomes $M^+$ and $M^-$, the later measurement provides no additional
information; that $S_x$ has opposite values for particles $a$ and $b$ is a
consequence of the initial state \eqref{eqn93}.

\xb
\outl{Discussion. EPR paradox = Paradox of 1-particle measurement}
\xa

We are now in a position to discuss the Bohm version of EPR using a consistent
theory of quantum measurements. The analysis based on the history family
\eqref{eqn97} indicates that one can, indeed, infer from a measurement of $S_z$
on particle $a$ the value of $S_z$ for particle $b$. However, see
\eqref{eqn101}, the $S_z$ measurement of particle $a$ tells one nothing about
$S_x$ for either particle $a$ or particle $b$. To which the response might be:
Make an $S_x$ measurement on particle $a$, and the outcome will then tell one
$S_x$ for particle $b$. This is entirely correct, but of course one cannot
measure both $S_x$ and $S_z$ for particle $a$, because there is nothing there
to be measured, at least if one is using Hilbert space quantum mechanics. As
for the counterfactual: ``$S_z$ was measured for particle $a$ and the value was
$+1/2$, but if instead $S_x$ had been measured its value would have been either
$+1/2$ or $-1/2$,'' this is blocked by the single framework rule applied to
quantum counterfactuals (Ch.~19 of \cite{Grff02c}). Thus the entire EPR
``paradox'' when analyzed from this point of view is nothing more than a
particular application of the ``paradox'' that in Hilbert space quantum
mechanics one cannot simultaneously assign values to $S_z$ and $S_x$ for a
spin-half particle. The issue is entirely a matter of what one can say about
the measurement of particle $a$. Particle $b$, together with entanglement, Bell
inequalities, possible nonlocality, etc., are from this perspective entirely
irrelevant. To be sure, entanglement, locality, and the like are in and of
themselves interesting topics; for a detailed discussion from the consistent histories point of
view, see \cite{Grff11b, Grff11} and Chs.~23 and 24 of \cite{Grff02c}.


\xb
\section{Conclusion \label{sct6}}
\xa

\xb
\outl{Virtues of solution to 2d measurement problem using CH}
\xa

We have shown that a satisfactory solution to the second measurement
problem---inferring a prior microscopic state of affairs from the macroscopic
outcome (pointer position) of a measurement described using quantum
principles---exists for a significant class of projective and generalized
(POVM) measurements. The approach using consistent histories is
mathematically sound, gives reasonable physical results, and does not lead to
paradoxes. Unlike current textbook treatments of measurements it makes no use
of ad hoc principles and special rules that apply only when measurements are
being made; instead the entire measurement process is analyzed using basic
quantum principles that apply to all physical processes, whether
microscopic or macroscopic.

\xb
\outl{Backwards map $Q^k=J\ad M^k J$ useful for finding property corresponding
  to  outcome $k$}
\xa

A useful feature of the approach used here is the backwards map $Q^k=J\ad M^k
J$, \eqref{eqn35}, relating a POVM element $Q^k$ to the projector $M^k$ on a
subspace that corresponds to outcome (``pointer position'') $k$. It is helpful
for identifying an earlier microscopic property or properties that resulted in
outcome $k$, even though it does not always give a precise answer. It is a
significant addition to, while at the same time completely consistent with, the
discussion of measurements in Chs.~27 and 28 of \cite{Grff02c}. And it would
seem to be particularly useful for analyzing weak measurements in terms of
physical properties rather than weak values, as illustrated by the simple
example in Sec.~\ref{sbct5.4}.

\xb
\outl{Applications in Sec.~\ref{sct5};  Qm
  noncontextuality and EPRB of particular significance}
\xa

The applications in Secs.~\ref{sbct5.1} to \ref{sbct5.4} are relatively simple
illustrations of the measurement formalism in Sec.~\ref{sct4}, but the last two
applications, Secs.~\ref{sbct5.5} and \ref{sbct5.6}, address issues about which
there is quite a bit of confusion in the published literature. Claims that
quantum mechanics is ``contextual'' are incorrect if that term is interpreted
in the sense introduced by Bell and used by Mermin. This has been pointed out
previously \cite{Grff13b}, but one may hope that the quite specific example
worked out in Sec.~\ref{sbct5.5} will result in a more precise definition of
the term ``contextual'' on the part those who claim that quantum mechanics is
contextual, or perhaps the withdrawal or modification of these claims. While
the nonexistence of nonlocal influences in Bohm's version of the
Einstein-Podolsky-Rosen paradox has been pointed out previously (see
\cite{Grff11} and the references given there), the analysis in
Sec.~\ref{sbct5.6} should help to further pin down the source of Bell's
mistake: he did not have a solution to the second measurement problem (or, for
that matter, the first, see \cite{Bll901}).

\xb
\outl{Key CH features: properties = subspaces; stochastic; histories;
  consistency}
\xa

It is worth listing the fundamental quantum principles which make the consistent histories
analysis possible. First, as we learned from von Neumann (Sec.~II.5 of
\cite{vNmn32b}), quantum properties (attributes of a physical system that can
be true or false) correspond to subspaces of the quantum Hilbert space: no need
for additional ``hidden variables.'' Next, following a proposal by Born
\cite{Brn26}, quantum time dependence is inherently stochastic: Schr\"odinger's
unitary time evolution should be used for calculating probabilities of events
rather than determining them. Stochastic quantum time development can be
described using histories represented by tensor products on a history Hilbert
space, as first pointed out by Isham \cite{Ishm94}. Assigning probabilities to
quantum histories of a closed system using the extended Born rule requires the
use of sample spaces satisfying consistency conditions---those used here are
the medium decoherence conditions of Gell-Mann and Hartle \cite{GMHr93}.

\xb
\outl{Absence of unicity, multiple descriptions + SFR, allow soln to 2d
  measurement problem}
\xa

Finally, a key principle that makes a clean break with classical thinking, and
hence is often misunderstood by critics of consistent histories, is the abandonment of what
elsewhere (Sec.~27.3 of \cite{Grff02c}) has been called the \emph{principle of
  unicity}: the idea that the universe, or at least that part of it which forms
a closed physical system, must at any given time be in a single, well-defined
physical state, a single point in a classical phase space. By contrast, the consistent histories
approach gives the physicist liberty to construct alternative quantum
descriptions---frameworks---which are incompatible with one another (and thus
cannot be combined, the single framework rule), each of which can make an equal
claim to describing some aspect of physical reality. That freedom, discussed in
greater detail in \cite{Grff14}, is important for resolving both the first and
the second measurement problem. As for the first problem, there is nothing
fundamentally wrong with using unitary time evolution leading to a
superposition state of different pointer positions, but this is of no use for
discussing the measurement as having specific outcomes. Once unicity has been
abandoned there is a perfectly good framework in which the pointer takes
well-defined positions, each with some probability. As for the second problem,
the textbook procedure that employs unitary evolution up to the time when the
particle begins to interact with the apparatus is perfectly good quantum
mechanics, but claiming that this is the \emph{only} valid quantum description
stands in the way of reaching the conclusion, using an appropriate framework,
that the apparatus constructed by a competent experimenter actually did measure
what it was designed to measure.

\xb
\outl{Need for a better understanding of `measurement' in textbooks}
\xa

A proper understanding of what it is that quantum measurements measure should
lead to a better physical understanding of the quantum world, and will, one
hopes, someday replace the unsatisfactory discussion of quantum principles
found in current textbooks. Students find introductory quantum theory hard to
understand both because the mathematics is unfamiliar and because its
connection with physical concepts seems obscure. They are not helped by the way
``measurement'' suddenly shows up in an almost magical way in textbook quantum
mechanics. Somehow it doesn't look like good physics. And it isn't. Students
deserve something better.


\begin{thebibliography}{10}

\bibitem{Vdmn13}
L.~Vaidman.
\newblock Past of a quantum particle.
\newblock {\em Phys. Rev. A}, 87:052104, 2013.

\bibitem{Grff16}
Robert~B. Griffiths.
\newblock Particle path through a nested {M}ach-{Z}ehnder interferometer.
\newblock {\em Phys. Rev. A}, 94:032115, 2016.
\newblock arXiv:1604.04596.

\bibitem{LAAZ15}
Zheng-Hong Li, M.~Al-Amri, and M.~Suhail Zubairy.
\newblock Direct counterfactual transmission of a quantum state.
\newblock {\em Phys. Rev. A}, 92:052315, 2015.

\bibitem{Vdmn16}
Lev Vaidman.
\newblock Comment on {``D}irect counterfactual transmission of a quantum
  state''.
\newblock {\em Phys. Rev. A}, 93:066301, 2016.
\newblock arXiv:1511.06615.

\bibitem{LAAZ16}
Zheng-Hong Li, M.~Al-Amri, and M.~Suhail Zubairy.
\newblock Reply to {``C}omment on {`D}irect counterfactual transmission of a
  quantum state'\ ''.
\newblock {\em Phys. Rev. A}, 93:066302, 2016.

\bibitem{Gsn17}
Nicolas Gisin.
\newblock Collapse. {W}hat else?
\newblock arXiv:1701.08300, 2017.

\bibitem{Wnbr15}
Steven Weinberg.
\newblock {\em Lectures on quantum mechanics}.
\newblock Cambridge University Press, Cambridge, U.\ K., 2d edition, 2015.

\bibitem{Grff02c}
Robert~B. Griffiths.
\newblock {\em Consistent Quantum Theory}.
\newblock Cambridge University Press, Cambridge, U.K., 2002.
\newblock http://quantum.phys.cmu.edu/CQT/.

\bibitem{Grff13b}
Robert~B. Griffiths.
\newblock Hilbert space quantum mechanics is noncontextual.
\newblock {\em Stud. Hist. Phil. Mod. Phys.}, 44:174--181, 2013.
\newblock arXiv:1201.1510.

\bibitem{Jmmr74}
Max Jammer.
\newblock {\em The Philosophy of Quantum Mechanics}.
\newblock Wiley, New York, 1974.

\bibitem{BcVl09}
Guido Bacciagaluppi and Antony Valentini.
\newblock {\em Quantum Theory at the Crossroads}.
\newblock Cambridge University Press, New York, 2009.
\newblock arXiv:quant-ph/0609184.

\bibitem{FyLS651}
R.~P. Feynman, R.~B. Leighton, and M.~Sands.
\newblock {\em The Feynman Lectures on Physics}, volume III: Quantum Mechanics.
\newblock Addison-Wesley, Reading, Mass., 1965.
\newblock Ch.~1.

\bibitem{Whlr78}
John~Archibald Wheeler.
\newblock The ``{P}ast'' and the ``{D}elayed-{C}hoice'' {D}ouble-{S}lit
  {E}xperiment.
\newblock In A.~R. Marlow, editor, {\em Mathematical Foundations of Quantum
  Theory}, pages 9--48. Academic Press, New York, 1978.

\bibitem{Grff14b}
Robert~B. Griffiths.
\newblock The {C}onsistent {H}istories {A}pproach to {Q}uantum {M}echanics.
\newblock {\em Stanford Encyclopedia of Philosophy}, 2014.
\newblock http://plato.stanford.edu/entries/qm-consistent-histories/.

\bibitem{Grff14}
Robert~B. Griffiths.
\newblock The {N}ew {Q}uantum {L}ogic.
\newblock {\em Found. Phys.}, 44:610--640, 2014.
\newblock arXiv:1311.2619.

\bibitem{vNmn32b}
Johann von Neumann.
\newblock {\em Mathematische Grundlagen der Quantenmechanik}.
\newblock Springer-Verlag, Berlin, 1932.
\newblock English translation by R. T. Beyer: \textit{Mathematical Foundations
  of Quantum Mechanics}, Princeton University Press, Princeton (1955).

\bibitem{BrvN36}
G.~Birkhoff and J.~von Neumann.
\newblock The logic of quantum mechanics.
\newblock {\em Ann. Math.}, 37:823--843, 1936.

\bibitem{Mttl09}
Peter Mittelstaedt.
\newblock Quantum logic.
\newblock In Daniel Greenberger, Klaus Hentschel, and Friedel Weinert, editors,
  {\em Compendium of Quantum Physics}, pages 601--604. Springer-Verlag, Berlin,
  2009.

\bibitem{Grff13}
Robert~B. Griffiths.
\newblock A consistent quantum ontology.
\newblock {\em Stud. Hist. Phil. Mod. Phys.}, 44:93--114, 2013.
\newblock arXiv:1105.3932.

\bibitem{Shr94}
Peter~W. Shor.
\newblock Algorithms for quantum computation: discrete logarithms and
  factoring.
\newblock In Shafi Goldwasser, editor, {\em Proceedings of the 35th Annual
  Symposium on Foundations of Computer Science}, pages 124--134. IEEE Computer
  Society Press, Los Alamitos, California, 1994.

\bibitem{Grjy05}
Edward Gerjuoy.
\newblock Shor's factoring algorithm and modern cryptography. {A}n illustration
  of the capabilities inherent in quantum computers.
\newblock {\em Am. J. Phys.}, 73:521--540, 2005.

\bibitem{Ldrs06}
Gerhart L{\"u}ders.
\newblock Concerning the state-change due to the measurement process.
\newblock {\em Ann. Phys. (Leipzig)}, 15:663--670, 2006.

\bibitem{BsLh09b}
Paul Busch and Pekka Lahti.
\newblock L{\"u}der's {R}ule.
\newblock In Daniel Greenberger, Klaus Hentschel, and Friedel Weinert, editors,
  {\em Compendium of Quantum Physics}, pages 356--358. Springer-Verlag, Berlin,
  2009.

\bibitem{TmCh13}
Boaz Tamir and Eliahu Cohen.
\newblock Introduction to weak measurements and weak values.
\newblock {\em Quanta}, 2:7--17, 2013.

\bibitem{Svns13}
Bengt E.~Y. Svensson.
\newblock Pedagogical review of quantum measurement theory with an emphasis on
  weak measurements.
\newblock {\em Quanta}, 2:18--49, 2013.
\newblock arXiv:1202.5148v.

\bibitem{Wkpd17}
Quantum tomography.
\newblock {\em Wikipedia}, 2017.
\newblock en.wikipedia.org/wiki/Quantum\_tomography.

\bibitem{Smao17}
Andrew~W. Simmons, Joel~J. Wallman, Hakop Pashayan, Stephen~D. Bartlett, and
  Terry Rudolph.
\newblock Contextuality under weak assumptions.
\newblock {\em New J. Phys.}, 19:033030, 2017.
\newblock arXiv:1610.06897.

\bibitem{Bll661}
John~S. Bell.
\newblock On the problem of hidden variables in quantum mechanics.
\newblock {\em Rev. Mod. Phys.}, 38:447--452, 1966.
\newblock Reprinted in John S. Bell, \textit{Speakable and Unspeakable in
  Quantum Mechanics, 2d ed.} (Cambridge University Press, 2004), pp.~1-13.

\bibitem{Mrmn93}
N.~David Mermin.
\newblock Hidden variables and the two theorems of {J}ohn {B}ell.
\newblock {\em Rev. Mod. Phys.}, 65:803--815, 1993.

\bibitem{Prs93}
Asher Peres.
\newblock {\em Quantum Theory: Concepts and Methods}.
\newblock Kluwer Academic Publishers, Dordrecht, The Netherlands, 1993.

\bibitem{Grff99}
Robert~B. Griffiths.
\newblock Consistent quantum counterfactuals.
\newblock {\em Phys. Rev. A}, 60:5--9, 1999.

\bibitem{Lloe01}
F.~Lalo{\"e}.
\newblock Do we really understand quantum mechanics? {S}trange correlations,
  paradoxes, and theorems.
\newblock {\em Am. J. Phys.}, 69:655--701, 2001.

\bibitem{Lloe12}
Franck Lalo{\"e}.
\newblock {\em Do We Really Understand Quantum Mechanics?}
\newblock Cambridge University Press, Cambridge, U. K., 2012.

\bibitem{Bll901}
J.~S. Bell.
\newblock Against measurement.
\newblock In Arthur~I. Miller, editor, {\em Sixty-Two Years of Uncertainty},
  pages 17--31. Plenum Press, New York, 1990.
\newblock Reprinted in John S. Bell, \textit{Speakable and Unspeakable in
  Quantum Mechanics, 2d ed.} (Cambridge University Press, 2004), pp.~213-231.

\bibitem{Glds12}
Sheldon Goldstein.
\newblock Bohmian mechanics.
\newblock {\em Stanford Encyclopedia of Philosophy}, 2012.
\newblock http://plato.stanford.edu/entries/qm-bohm/.

\bibitem{Fne82}
Arthur Fine.
\newblock Joint distributions, quantum correlations, and commuting observables.
\newblock {\em J. Math. Phys.}, 23:1306--1310, 1982.

\bibitem{Grff11}
Robert~B. Griffiths.
\newblock Quantum locality.
\newblock {\em Found. Phys.}, 41:705--733, 2011.
\newblock arXiv:0908.2914.

\bibitem{Spkk07}
Robert~W. Spekkens.
\newblock Evidence for the epistemic view of quantum states: {A} toy theory.
\newblock {\em Phys. Rev. A}, 75:032110, 2007.
\newblock arXiv:quant-ph/0401052v2.

\bibitem{Lfr14}
M.~S. Leifer.
\newblock Is the quantum state real? a review of $\psi$-ontology theorems.
\newblock {\em Quanta}, 3:67--155, 2014.
\newblock arXiv:1409.1570.

\bibitem{JnLf15}
David Jennings and Matthew Leifer.
\newblock No return to classical reality.
\newblock {\em Contemporary Physics}, 56:60--82, 2016.
\newblock arXiv:1501.03202.

\bibitem{KcSp67}
Simon Kochen and E.~P. Specker.
\newblock The problem of hidden variables in quantum mechanics.
\newblock {\em J. Math. Mech.}, 17:59--87, 1967.

\bibitem{AbBM17}
Samson Abramsky, Rui~Soares Barbosa, and Shane Mansfield.
\newblock The contextual fraction as a measure of contextuality.
\newblock {\em Phys. Rev. Lett.}, 119:050504, 2017.
\newblock arXiv:1705.07918.

\bibitem{EnPr35}
A.~Einstein, B.~Podolsky, and N.~Rosen.
\newblock Can quantum-mechanical description of physical reality be considered
  complete?
\newblock {\em Phys. Rev.}, 47:777--780, 1935.

\bibitem{Bhm51s}
David Bohm.
\newblock {\em Quantum Theory}, chapter~22.
\newblock Prentice Hall, Englewood Cliffs, N.J., 1951.

\bibitem{Grff11b}
Robert~B. Griffiths.
\newblock E{P}{R}, {B}ell, and quantum locality.
\newblock {\em Am. J. Phys.}, 79:954--965, 2011.
\newblock arXiv:1007.4281.

\bibitem{Brn26}
Max Born.
\newblock Zur {Q}uantenmechanik der {S}to\ss vorg{\"a}nge.
\newblock {\em Z. Phys.}, 37:863--867, 1926.

\bibitem{Ishm94}
C.~J. Isham.
\newblock Quantum logic and the histories approach to quantum theory.
\newblock {\em J. Math. Phys.}, 35:2157--2185, 1994.

\bibitem{GMHr93}
Murray Gell-Mann and James~B. Hartle.
\newblock Classical equations for quantum systems.
\newblock {\em Phys. Rev. D}, 47:3345--3382, 1993.

\end{thebibliography}

\xb
\end{document}